\documentclass{aa}  

\usepackage{graphicx}
\usepackage{txfonts}
\begin{document}

   \title{Realising efficient computation of individual frequencies for \\ red-giant models}

   \subtitle{The Truncated Scanning Method}

   \author{J. R. Larsen\inst{1}\fnmsep\thanks{E-mail: jensrl@phys.au.dk}
          \and
          J. Christensen-Dalsgaard\inst{1}
          \and 
          M. S. Lundkvist\inst{1}
          \and  
          J. L. Rørsted \inst{1,2}
          \and
          M. L. Winther \inst{1}
          \and
          H. Kjeldsen\inst{1,2}
          }

   \institute{Stellar Astrophysics Centre (SAC), Department of Physics and Astronomy, Aarhus University,
              Ny Munkegade 120, 8000 Aarhus C, Denmark \\
              \and
             Aarhus Space Centre (SpaCe), Department of Physics and Astronomy, Aarhus University, Ny Munkegade 120, 8000              Aarhus C, Denmark
            }
   \date{Received 11/01/2024; Accepted 27/08/2024}

 
  \abstract
   {In order to improve the asteroseismic modelling efforts for red-giant stars, the numerical computation of theoretical individual oscillation modes for evolved red-giant models has to be made feasible.}
   {We aim to derive a method for circumventing the computational cost of computing oscillation spectra for models of red-giant stars with an average large frequency separation $\Delta\nu<15$ $\mu$Hz, thereby allowing for asteroseismic investigations of giants utilising individual frequencies.}
   {The proposed Truncated Scanning Method serves as a novel method detailing how the observable individual $l=0,1,2$ frequencies of red giants may be computed on realistic timescales through so-called \emph{model truncation}. By carefully removing the innermost region of the stellar models, the g-mode influence on the oscillation spectra may be avoided, allowing estimation of the observable regions from the resulting pure p-mode oscillations. The appropriate observable frequency regions are subsequently scanned for the complete and un-truncated stellar model. The observable regions are determined by considering the limitations on observability from the internal mode coupling and damping, yielding consistent frequency spectra obtained at a much reduced computational cost.}
   {The Truncated Scanning Method proves the feasibility of obtaining the individual frequencies of red-giant models for a wide range of applications and research, demonstrating an improved computational efficiency by a factor of 10 or better. This means that the inclusion of $l=1,2$ individual frequencies is now a possibility in future asteroseismic modelling efforts of red-giant stars. Further potential avenues for improvements to this method are outlined for future pursuits.}
   {}


   \maketitle
%

\section{Introduction}
Asteroseismology is the study of stellar oscillations, which allows for detailed inferences regarding the interior structure of stars; a property that conventional techniques such as photometry and spectroscopy fail to obtain due to the obscuring stellar opacity. In-depth analyses using individual frequencies are routinely done for the Sun and other solar-type stars along the main sequence (MS) -- see e.g. \citet{Houdek09}; \citet{Aguirre17}; \citet{Winther23}. Here the stars behave according to the predictions of the asymptotic theory (see Sect.~\ref{sec:RedGiantOsc}) and are stochastically excited by convection -- we say that they exhibit \emph{solar-like} oscillations. As the stars evolve past the MS turn-off and onto the sub-giant branch (SGB), the utilisation of individual oscillations frequencies can still be employed. Here, the stars begin to develop features deviating from the asymptotic theory, but to a restricted extent such that it can be handled accordingly by our oscillation codes. The deviations can provide sensitive information on the interior structure, yielding constraints on the characterisation of the SGB stars (e.g., \citealt{CD_etaboo95}; \citealt{Metcalfe_10}; \citealt{Grundahl17}; \citealt{Stokholm19}). Yet, when wishing to extend the considerations to the more evolved counterparts, the red giants, our efforts are obstructed by the nature of their oscillation spectra. 

Asteroseismology and the inferences possible from the individual frequencies could provide valuable insights into the nature of red giants -- aiding in their characterisation and understanding. Yet, the crucial criteria for making such inferences is the possibility to link observed oscillations to the oscillation modes of the stellar models. On the observational side, the asteroseismic data available for giants have been revolutionised by space missions such as CoRoT (\citealt{Baglin06}) and \textit{Kepler} (\citealt{Gilliland10}). Deriving the observed asteroseismic parameters for the expansive collection of available timeseries has been done extensively, and as the oscillation amplitudes roughly scale as the ratio between the stellar luminosity and mass, $L/M$, as summarised by \citet{Kjeldsen95}, evolved stars have readily apparent oscillation patterns. Contrarily, the oscillatory properties of the red-giant stellar models impose the constraints to our investigations \citep[e.g.,][]{GiantStarSeis}.  

A stellar model represents a snapshot in time of the interior stellar profile produced by a stellar evolution code. Various resulting stellar tracks and isochrone compilations exist (e.g. \citealt{Pietrinferni04}; \citealt{Dotter08}; \citealt{Dotter16}; \citealt{Hidalgo18}), and the modelling of stars on the red-giant branch (RGB) hinges on connecting such compilations to observations \citep{Salaris2002}. As a part of the Aarhus red-giant challenge, an evaluation of the numerical accuracy and consistency of an ensemble of evolution codes used for red-giant modelling was carried out \citep{RGCI}. The connection between asteroseismology and the stellar evolution codes with the stellar structures they produce, however, is far from a trivial problem. The subsequent paper by \citet{RGCII} discussed the complex relation between evolved stellar structures and oscillation properties, employing the same oscillation code as the present work. 

Previous asteroseismic characterisation and modelling of red giants has therefore relied on the global parameters  (\citealt{Mosser11}; \citealt{Hekker11}; \citealt{Wang23} and references therein). These are the large frequency separation $\Delta\nu$, the frequency of maximum oscillatory power $\nu_\mathrm{max}$ and the dipolar period spacing $\Delta\Pi_1$. In practice, $\Delta\nu$ and $\nu_\mathrm{max}$ are obtainable from the scaling relations \citep{Kjeldsen95} (see Eqs.~\ref{eq:DnuScal} and \ref{eq:NumaxScal}), making their retrieval readily available for the stellar models for comparison with observed values. We note that in recent years some concern has arisen about the applicability of the scaling relations, particularly for $\nu_\mathrm{max}$, for evolved stars with sub-solar metallicity (\citealt{Epstein14}; \citealt{Viani17}). Notably, $\Delta\Pi_1$ has been successfully used as a constraint on the stellar interior, having been shown by \citet{Bedding11} to clearly distinguish helium-burning red-clump and first-ascent red-giant stars. 

In the effort to constrain the stellar interior using asteroseismology, it is possible to go one step further than the period spacing through a consideration of the gravity phase offset $\epsilon_\mathrm{g}$ in the asymptotic relation for gravity modes. The $\epsilon_\mathrm{g}$ term can assist in constraining the determination of $\Delta\Pi_1$ \citep{Buysschaert16} as well as identifying and studying the borders of the internal cavity for the gravity modes \citep{Pincon19}. Further diagnostic power is provided by the coupling factor $q$ of mixed modes (see Sect.~\ref{sec:RedGiantOsc}) allowing for insights into the intermediate regions between the stellar envelope and core. The coupling factor has also been shown to vary with stellar mass and evolutionary stage \citep{Mosser17}. 

Fitting the red-giant models directly to the entire spectrum of observed individual frequencies has so far been unfeasible, leaving the potential of such investigations unexplored. However, the properties of the individual red-giant oscillations provides a unique gateway for observational constraints throughout the stellar interior. Realising this link between the observed and modelled individual frequencies by circumventing the computational cost of obtaining the oscillations for red-giant models is the focus of this paper. 

In the following we outline the oscillatory behaviour of red giants in Sect.~\ref{sec:RedGiantOsc}, emphasising the features they develop which both complicate the numerical computations yet also offers constraints for the stellar interior profiles, before discussing the observability of the pulsations in real stars. The necessary considerations and requirements for computing giant-star oscillation spectra are then introduced in Sect.~\ref{sec:CalculatingGiantOsc}. The fundamental idea and approach of this work in realising the computations is outlined in Sect.~\ref{sec:TruncatingModels}, along with a brief overview of the complications encountered. Retrieving proper observable frequency intervals for a given stellar model is covered in Sect.~\ref{sec:ScanningIntervals}, before validating the method on a stellar grid covering a wide parameter space in Sect.~\ref{sec:Validation}. Section~\ref{sec:Results} presents the resulting outcomes of the \emph{Truncated Scanning Method} for a representative model. Lastly, in Sect.~\ref{sec:DiscussionAndConclusion} we discuss further avenues for improvement of the concepts and methods presented, before conclusions on the work are made.

\section{Oscillations of red giants}\label{sec:RedGiantOsc}
In order to understand the reason behind the troublesome oscillation properties of red-giant models, we return to the fundamental asymptotic description for stellar oscillations. Under the assumption of spherical symmetry, neglecting the effects of rotation, the stellar oscillation modes may be described at co-latitude $\theta$ and longitude $\phi$ by spherical harmonics, $Y^m_l(\theta,\phi)$. In interpreting the spherical harmonics we may describe the modes by their mode degree $l$ as the total number of nodes on the stellar surface, and the azimuthal order $m$ counting the number of nodal lines crossing the equator. In spherically symmetric stars the pulsations become independent of $m$. The final addition for mode characterisation is the radial order $n$, defining the number and properties of nodes in the radial direction \citep[see, e.g.,][]{AsteroACDK}. For purely acoustic modes, or acoustic resonances for mixed modes (see Fig.~\ref{fig:InertiaPlot} and discussion thereof), this results in the asymptotic relation, in the first-order approximation, describing a given oscillation frequency of a given spherical degree $l$ and acoustic (pressure) mode order $n_{\rm p}$ as 
\begin{equation}\label{eq:AsympRel}
    \nu_{nl} \simeq \Delta\nu \left( n_{\rm p}+\frac{l}{2}+\epsilon \right) .
\end{equation}
Here, $\Delta\nu$ is the aforementioned large frequency separation describing the separation between modes of identical degree $l$ but consecutive order $n_{\rm p}$, and $\epsilon$ is a constant close to unity called the phase term \citep{White11}. 

\begin{figure}
    \resizebox{\hsize}{!}{\includegraphics[width=\linewidth]{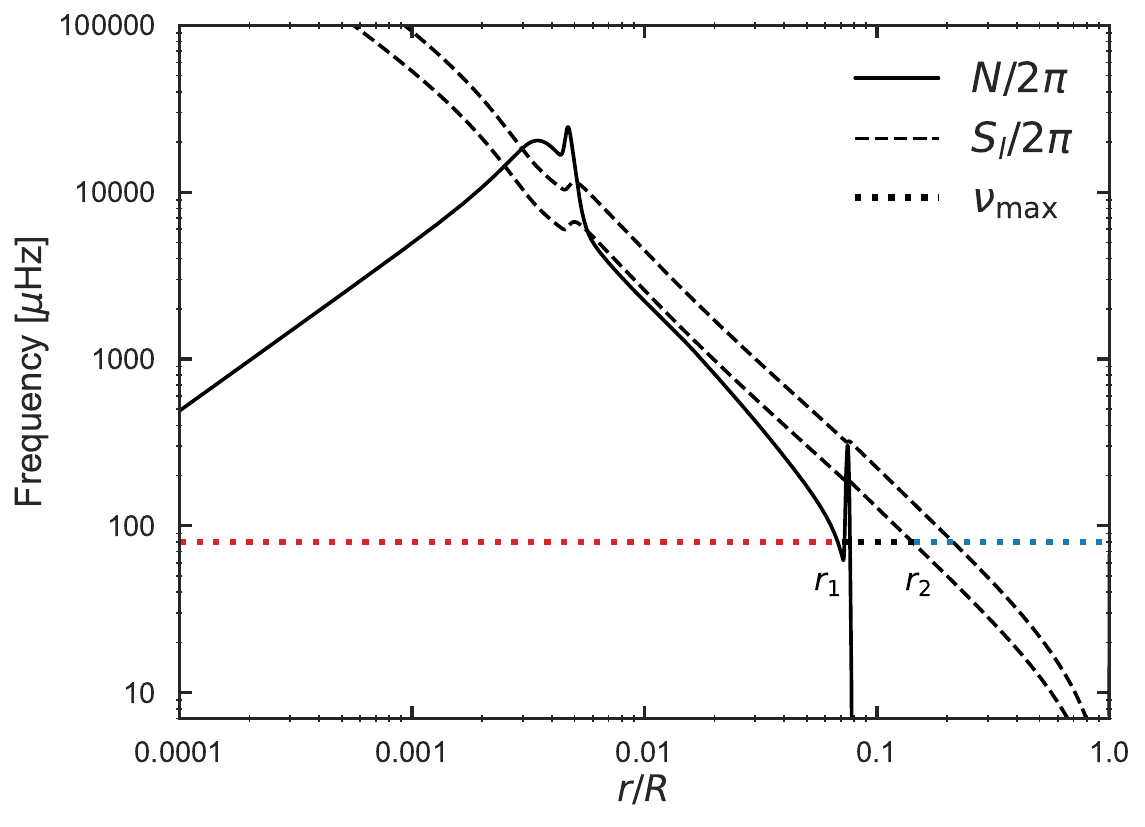}}
    \centering
    \caption{Characteristic frequencies for model $\mathcal{M}$ with parameters given in Table~\ref{tab:ModelM}, expressed in terms of cyclic frequencies $N/2\pi$ (solid), $S_1/2\pi$ and $S_2/2\pi$ (both dashed, $S_2$ residing further from the interior). The horizontal dotted line indicates the frequency of maximum power $\nu_\mathrm{max}$. The gravity-mode cavity is indicated by the red section with its upper boundary at $r_1$, and the pressure-mode cavity by the blue (with the lower boundary at $r_2$) for $l=1$. A prominent buoyancy glitch can be seen at $r/R\approx 0.075$.}
    \label{fig:CharFreqs}
\end{figure}
\begin{table}
\centering
\caption{The mass, radius, metallicity, large frequency separation and frequency of maximum oscillatory power of the evolved solar model referred to as model $\mathcal{M}$.}
\begin{tabular}{ccc}
\hline
\multicolumn{1}{c}{\thead{Stellar parameter}} & \multicolumn{1}{c}{\thead{Value}} \\ \hline 
Mass                            &  1.0 M$_\odot$ \\
Radius                          &  6.59 R$_\odot$ \\
$[\mathrm{Fe}/\mathrm{H}]$      &  0.0 dex         \\
$\Delta\nu$                     &  8.02 $\mu$Hz       \\
$\nu_\mathrm{max}$              &  32.4  $\mu$Hz        \\
\hline
\end{tabular}
\label{tab:ModelM}
\end{table}
Red-giant stars still exhibit solar-like oscillations, yet display additional features in their oscillation patterns \citep[for extensive reviews see, e.g.,][]{Chaplin13,GiantStarSeis}. This occurs as red-giant stars have evolved to form a dense degenerate core, resulting in the local gravitational acceleration, $g$, reaching very high values in the deep interior. By extension, so will the buoyancy frequency $N$ governing the propagation of the gravity modes defined as (\citealt{AsteroACDK}, Eq. 3.73)
\begin{equation}\label{eq:BuoyancyFreq}
   N^2=g\left(\frac{1}{\Gamma_1}\frac{\mathrm{d} \ln{p}}{\mathrm{d}r}-\frac{\mathrm{d} \ln{\rho}}{\mathrm{d}r} \right).
\end{equation}
Here, $r$ is the radial distance from the centre of the star, $p$ and $\rho$ denote pressure and density, respectively, and $\Gamma_1=(\partial \ln{p}/\partial \ln{\rho})_{\mathrm{ad}}$ is the first adiabatic exponent. In convectively stable regions $N^2$ is positive and g-modes can exist at frequencies $0<\nu\leq N^2$ -- within their trapping region. Meanwhile, as the radius of the star has increased, this, with a contribution from the decreasing temperature in the exterior, results in a reduction in the sound speed $c$, which leads to a fall in the Lamb frequency $S_l$ \citep{Lamb32}, defined as
\begin{equation}\label{eq:LambFreq}
    S_{l}^2=\frac{l(l+1)c^2}{r^2}.
\end{equation}
This means that we can describe the trapping region/acoustic cavity of a given p-mode as being between the characteristic acoustic frequency and acoustic cut-off frequency -- which governs the wave-behaviour near the surface layers -- as  $S_l\leq\nu_{nl}\leq\nu_{\mathrm{ac}}$. Figure~\ref{fig:CharFreqs} illustrates the characteristic frequencies as cyclic frequencies $N/2\pi$ and $S_l/2\pi$ within the interior of an evolved solar stellar model; the parameters of which is presented in Table~\ref{tab:ModelM}. This model is used to generate the representative figures throughout this work, and will be referred to simply as model $\mathcal{M}$.

The rise in $N$ and fall in $S_l$ for giant stars ultimately results in the g- and p-mode cavities existing in identical frequency ranges near the observable region in frequency, near $\nu_\mathrm{max}$, such as seen in Fig.~\ref{fig:CharFreqs}. This fact leads to the phenomenon of \emph{mixed-mode} oscillations once the separation between the cavities, the \emph{evanescent} region (marked by $r_1$ and $r_2$), become sufficiently small for a coupling between the outer acoustic modes and inner gravity modes to occur \citep{Jiang14}. In a sense, the g-modes can now leave an imprint on the p-modes which in turn propagates to the exterior of the star. Mixed modes exist for all non-radial modes in giants, though only the dipole and quadrupole modes have observable amplitudes in disk-integrated light. The radial modes with $l=0$ are unable to display mixed-mode characteristics, as no gravity modes of $l=0$ exist for the radial p-modes to couple to. The radial modes of giants thus display purely acoustic behaviour, and are simple in their numerical computation. In the subsequent work they are therefore trivially determined from the complete stellar models without the need for further treatment. 

\begin{figure}
    \resizebox{\hsize}{!}{\includegraphics[width=\linewidth]{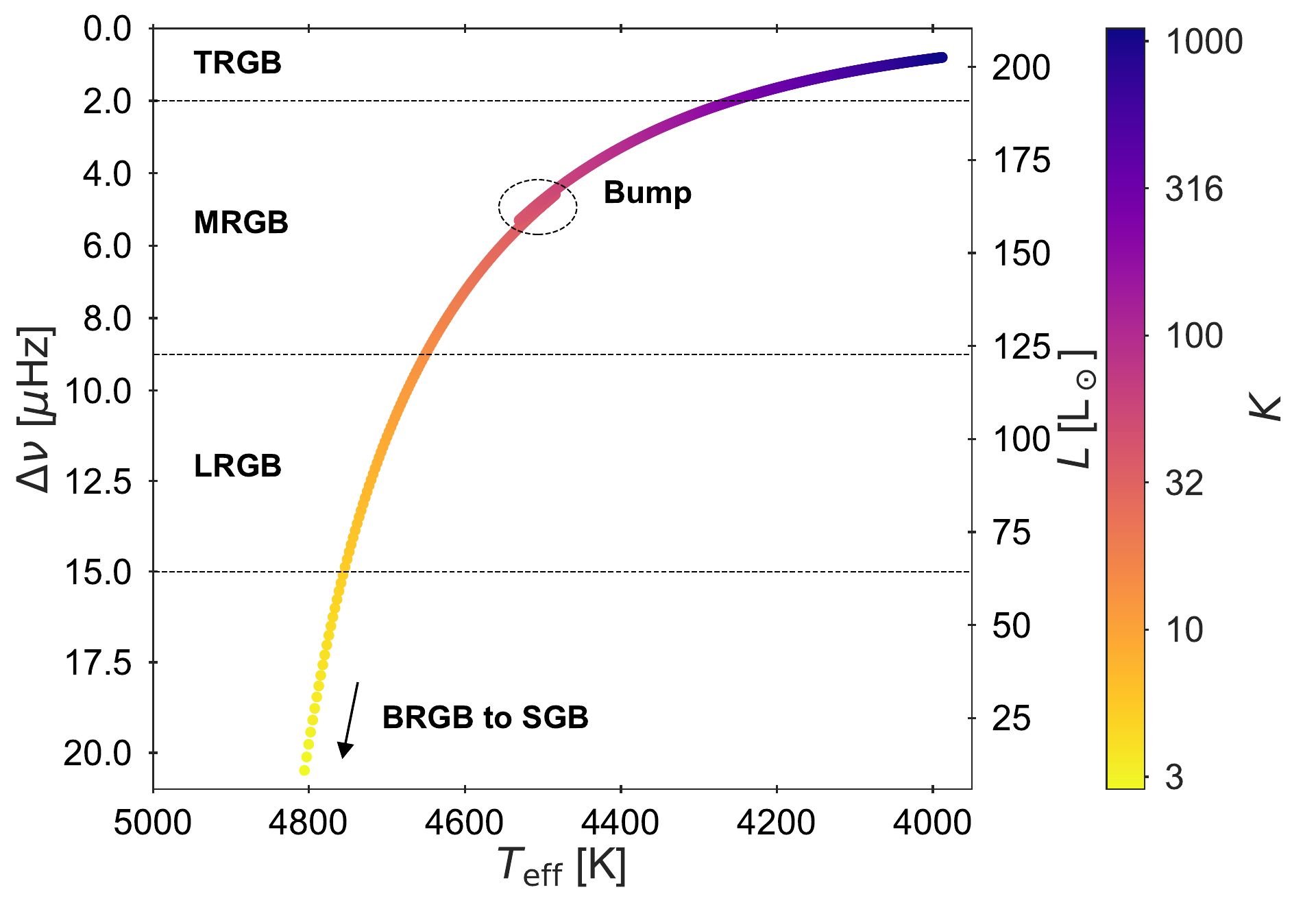}}
    \centering
    \caption{Evolutionary diagram for a solar stellar track with $M=1 \ M_\odot$ and $[\mathrm{Fe}/\mathrm{H}] = 0$ dex, plotting the large frequency separation $\Delta\nu$ and luminosity $L$ as a function of effective temperature $T_\mathrm{eff}$. The colour bar indicates the number of dipole modes in a $\Delta\nu$-wide frequency range around $\nu_{\rm max}$, according to Eq.~\ref{eq:NumModes}. The three classifications from top to bottom denote the tip (TRGB), middle (MRGB) and lower RGB (LRGB). }
    \label{fig:HRdiagram}
\end{figure}
The existence of the mixed modes for non-radial pulsations is the source of the complication we wish to circumvent. A single exterior pressure mode with acoustic mode order $n_{\rm p}$ may couple to a multitude of interior gravity modes, each with an associated mode order $n_{\rm g}$. This leads to dense oscillation spectra for giants with thousands of theoretical oscillation modes potentially being excited, which is computationally expensive to evaluate for each individual stellar model. However, the mixed modes display a sensitivity both to the outer stellar layers through their p-mode component, but also to the internal structure by the influence from the g-mode counterpart, leading to valuable information in modelling efforts (\citealt{Hjørringaard17}; \citealt{Mosser18}). 

For the purpose of characterising the effect of the evolutionary stage on the oscillatory nature along the RGB, the global parameters $\Delta\nu$, $\nu_\mathrm{max}$ and $\Delta\Pi_1$ can provide valuable insights. The number of possible theoretical stellar oscillations becomes a challenge for stars evolved beyond $\Delta\nu\approx15 \ \mu$Hz -- i.e. well into the RGB evolution, but prior to the location of the RGB luminosity bump \citep{Khan18}. This can be seen from the parameter $K$ that estimates the number of calculations that must be carried out for each $\Delta\nu$-wide frequency range in an RGB model to obtain all theoretical dipole modes \citep[Eq. 22 of][]{Mosser15},
\begin{equation}\label{eq:NumModes}
    K=\frac{\Delta\nu}{\Delta\Pi_1 \nu_{\mathrm{max}}^2}.
\end{equation}
Figure~\ref{fig:HRdiagram} illustrates this fact for the solar stellar track of model $\mathcal{M}$, covering an evolutionary region from the base of the RGB (BRGB) and onward, colour-coded by the value of $K$. The figure shows three classification regions of the RGB; the lower RGB (LRGB), middle RGB (MRGB) and the tip of the RGB (TRGB). The LRGB represents the initial region of interest, when the method proposed here starts to become relevant, the MRGB comprises the largest region of interest for the method and contains the RGB bump, while the TRGB is the last and most densely population region in terms of oscillation modes. The total number of calculations required becomes even larger with the inclusion of the quadrupole ($l=2$) modes and their period spacing $\Delta\Pi_2$. In this context, Fig.~\ref{fig:HRdiagram} and $K$ illustrates the rapidly increasing computational difficulties along the RGB.

Our method demonstrates a consistent way of isolating the frequency regions within the dense oscillation spectra of giants that are observable. This reduces the computation time substantially by allowing for a large fraction of the theoretical oscillations in each stellar model to be disregarded entirely. To isolate the most observable modes we consider the mode inertia
\begin{equation}\label{eq:ModeInertia}
    E=\frac{\int_V \rho |\delta \Vec{r}|^2 \mathrm{d}V}{M|\delta \Vec{r}|^2_{\textup{phot}}} .
\end{equation}
Here $\delta\Vec{r}$ denotes the displacement vector and "phot" the photospheric value of the vector, where the integral is performed over the volume $V$ of the star. An intuitive interpretation of the mode inertia is that the lower the inertia of a mode, the easier it is excited into oscillation, as a smaller fraction of the star's mass is involved in the pulsation \citep{Chaplin13}. The approach is therefore to isolate the mixed modes of each acoustic mode order $n_{\rm p}$ carrying the lowest inertia for a given stellar model. Crucially, the identification of these mixed modes must be realised without the need for calculating the entire pulsation spectrum. This approach and Eq.~\ref{eq:ModeInertia} can be represented in a plot of the mode inertia, such as displayed in Fig.~\ref{fig:InertiaPlot}. It is apparent that the radial modes do not display mixed-mode nature. For non-radial modes the inertia shows \emph{acoustic resonances}, characterised by an acoustic mode order $n_{\rm p}$ at frequencies approximately satisfying Eq.~\ref{eq:AsympRel}. Isolating the most observable mixed modes thus entails identifying the modes in these "inertia valleys", while excluding the many modes exhibiting higher inertia seen at the peaks. 
\begin{figure}
    \resizebox{\hsize}{!}{\includegraphics[width=\linewidth]{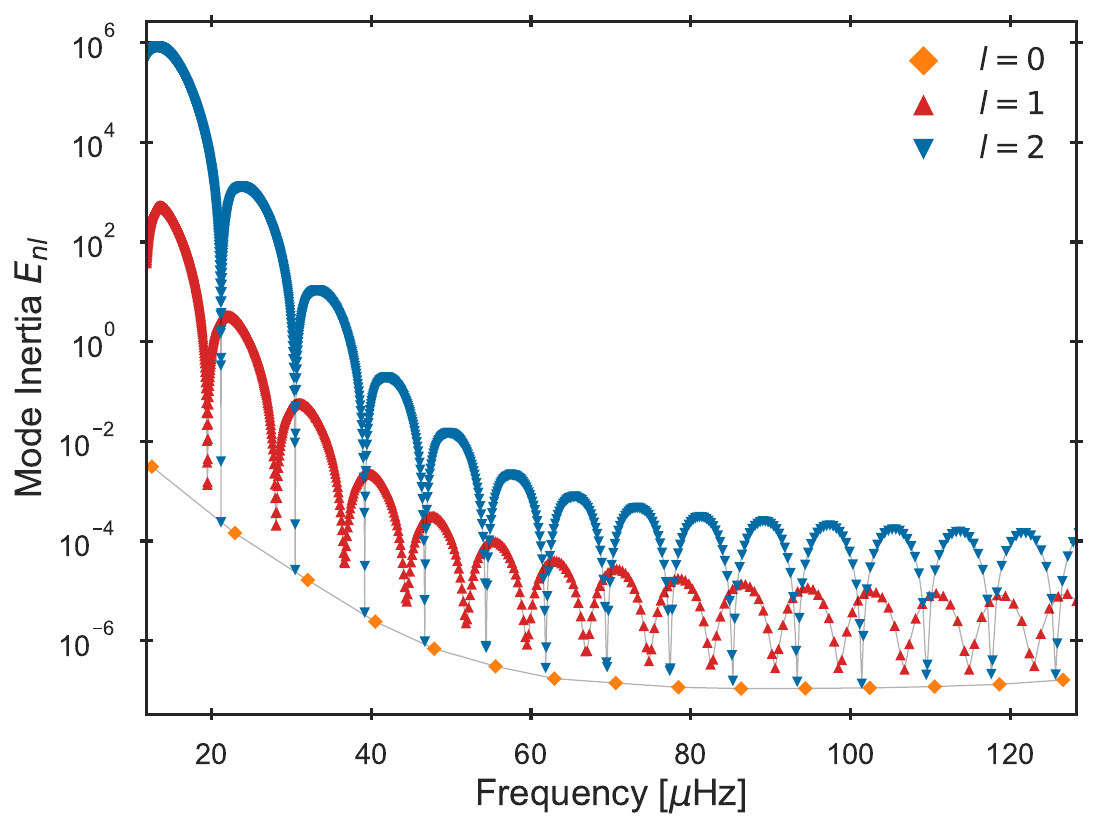}}
    \centering
    \caption{Inertia plot of model $\mathcal{M}$. The red upwards triangles and blue downwards triangles denote the dipole and quadrupole modes, respectively. The radial modes are depicted as the orange diamonds. The plotted modes are connected by solid lines to guide the eye.}
    \label{fig:InertiaPlot}
\end{figure}

In characterising the stellar models considered during the development of this work, the large frequency separation and frequency of maximum oscillatory power were consistently used. Both parameters were obtained under the assumption of an isothermal atmosphere \citep{Chaplin13} such that  
\begin{align}
    \Delta\nu & \simeq   \left(\frac{M}{M_\odot}\right)^{0.5} \left(\frac{T_{\mathrm{eff}}}{T_{\mathrm{eff},\odot}} \right)^{3.0} \left(\frac{L}{L_\odot}\right)^{0.75}\Delta\nu_\odot , \label{eq:DnuScal}\\
    \nu_\mathrm{max} & \simeq  \left(\frac{M}{M_\odot} \right) \left(\frac{T_{\mathrm{eff}}}{T_{\mathrm{eff},\odot}} \right)^{3.5} \left(\frac{L}{L_\odot} \right)\nu_{\mathrm{max},\odot} , \label{eq:NumaxScal}
\end{align}
where the stellar parameters carry their usual interpretation. This formulation is used throughout this work whenever $\Delta\nu$ or $\nu_\mathrm{max}$ are referenced.

\subsection{Observability of red-giant oscillations}\label{subsec:Observability}
The previous section outlined the nature of red-giant oscillations and the situation we face in the stellar models. We now wish to clarify the phenomena related to the observability of the oscillations. Investigating such aspects has been done from both a theoretical and observational perspective in the past, offering insight into potential constraints to impose on our method. 

From the theoretical side, \citet{Dupret09} investigated three RGB models in detail to examine their oscillation spectra. They found that non-radial modes should be observable throughout the RGB, but becoming limited to the purely exterior p-modes beyond the RGB bump. The main cause of this is the degree of radiative damping affecting the lifetime of the oscillation, which becomes increasingly significant along the RGB. This interpretation was seconded by \citet{Grosjean14}, similarly finding that mixed-mode oscillations become undetectable for evolved red giants - where they argue that it is both an effect of the increased radiative damping as well as decrease in the coupling strength $q$. 

Observations from \emph{Kepler} were interpreted by \citet{Bedding10} and later by \citet{Stello14}. They confirm the above theoretical view, observing non-radial mixed modes for less evolved red giants. Yet as the star becomes more evolved, the mixed modes gradually disappear. However, \citet{Stello14} still recover the acoustic resonances of non-radial oscillations for stars with $\Delta\nu \lesssim 1 \ \mu$Hz when using the entire 4-year \emph{Kepler} timeseries, indicating the importance and application of asteroseismology all along the RGB (see their Fig.~4 for reference). 

To incorporate these considerations and constraints we follow the approach by \citet{Mosser18}, who utilise the properties of the $\zeta$-function. The $\zeta$-function is defined as the fraction of mode inertia from the interior relative to the total mode inertia, and can be reformulated to depend on $K$ and the mode coupling $q$ \citep{GiantStarSeis}. \citet{Mosser18} use the full width at half maximum of the $\zeta$-function to derive a formula for the observable region of mixed modes around an acoustic resonance, which expressed in frequency becomes 
\begin{equation}\label{eq:ObsWidth}
    \delta \nu_\mathrm{obs}(\nu) = \frac{2q(\nu)}{\pi} \sqrt{1+\frac{1}{Kq(\nu)}}, 
\end{equation}
where $q(\nu)$ is the coupling constant of a given frequency and $K$ being defined by Eq.~\ref{eq:NumModes} for the respective period spacing of the mode. Estimating $q$ for any given frequency within a stellar model was described by \citet{Shibahashi79},
\begin{equation}\label{eq:CouplingConstant}
    q = \frac{1}{4}\exp{\left(-2\int_{r_1}^{r_2}|\kappa|^{1/2} dr\right)}.
\end{equation}
Here, $r_1$ and $r_2$ denote the lower and upper radial coordinate of the evanescent region's extent, as indicated in Fig.~\ref{fig:CharFreqs}. The radial wave vector $\kappa$ is calculated following \citet{Shibahashi79} as 
\begin{equation}\label{eq:RadialWaveVector}
    \kappa = \frac{\omega^2}{c^2}\left( \frac{S_l^2}{\omega^2} -1 \right) \left( \frac{N^2}{\omega^2}-1\right),
\end{equation}
where $\omega=2\pi \nu$ is the angular frequency, $c$ is the sound speed in the deep interior and the characteristic frequencies $N$ and $S_l$ are defined in Eqs.~\ref{eq:BuoyancyFreq} and \ref{eq:LambFreq}, respectively. The expression for $q$ in Eq.~\ref{eq:CouplingConstant} has, by construction, an upper limit at $1/4$. This effectively results in a limitation on the dimensions of $q$ when used in Eq.~\ref{eq:ObsWidth}. 

The use of Eq.~\ref{eq:ObsWidth} enables us to overlay realistic constraints on the observability of the model frequencies retrieved. Its use and application within this method will be further outlined in Sect.~\ref{sec:ScanningIntervals}.

\begin{figure*}[t]
    \includegraphics[scale=0.56]{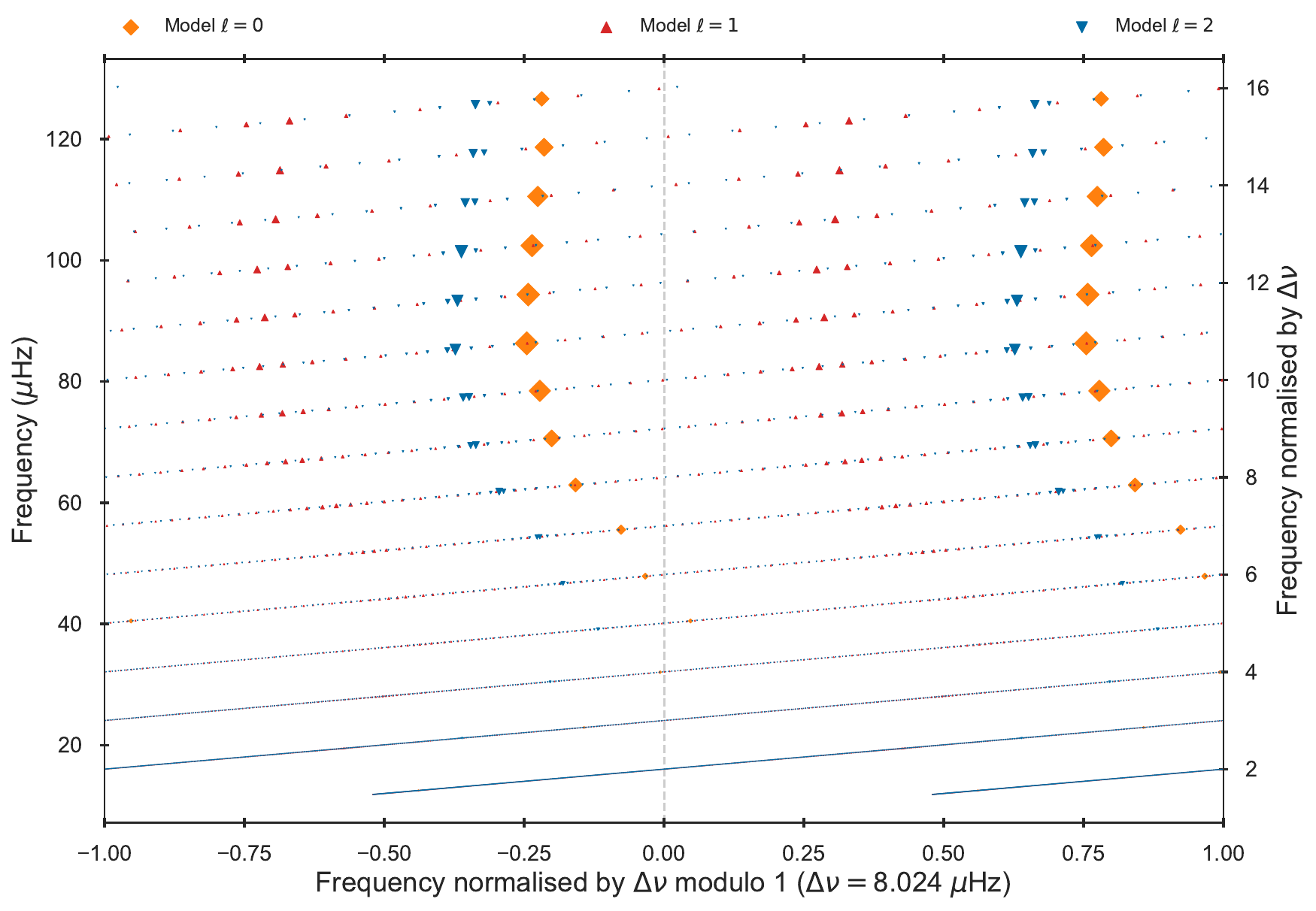}
    \centering
    \vspace{-2mm}
    \caption{Replicated {\'e}chelle diagram of a full calculation of model $\mathcal{M}$. The orange diamonds, red upwards triangles and blue downwards triangles depict the $l=0$, $l=1$ and $l=2$ modes, respectively for a model below the RGB bump. For each plotted mode, a scaling in size related to the inverse of its inertia is applied.}
    \label{fig:FullModEchelle}
    \vspace{-3mm}
\end{figure*}

\subsection{Numerical scheme}\label{subsec:ADIPLSNumScheme}
The oscillation code employed in this work is the \textit{Aarhus Adiabatic Pulsation Package} (ADIPLS; \citealt{ADIPLS08}). The code solves the adiabatic oscillation equations separately from the centre (or a truncation point; see Sect.~\ref{sec:TruncatingModels}) and the surface, identifying eigenfrequencies by requiring a continuous match of the two solutions. For radial modes, or in truncated models, the matching point is close to the surface of the model, while for nonradial oscillations in complete models it is typically located at the innermost maximum in the buoyancy frequency. The eigenfrequencies are determined by scanning in frequency and locating zeros in a matching determinant. The step in the scan is roughly based on the expected asymptotic behaviour of the frequencies. For truncated models we solve the second-order set of equations in the Cowling approximation, neglecting the perturbation to the gravitational potential, and obtain the frequency correction resulting from the potential perturbation through a perturbation analysis (\citealt{JCD82}, Eq. D1); for complete models the full set of equations is solved.
The equations are solved with a fourth-order Runge-Kutta technique \citep{Cash80}, using linear interpolation to obtain model values at points intermediate between the meshpoints.

In complete models, the central boundary conditions are obtained from an expansion to second order around the singular central point. In truncated models the truncation point is assumed to be in the evanescent region between the p- and g-mode trapping regions (see also Appendix \ref{app:TruncAlg}) and the inner boundary condition is set to obtain the solution decreasing exponentially towards the interior. The dynamic surface boundary condition assumes a continuous match to an isothermal atmosphere.

\section{Calculating giant-star oscillations}\label{sec:CalculatingGiantOsc}
Numerically calculating accurate oscillations for red-giant models that reflect the true pulsations of real stars involves changing the computational approach considerably. Here the modifications were made to ADIPLS, but the method could be extended to other pulsation codes. The process of implementing the necessary changes to a pulsation code in order to achieve the appropriate settings for complete red giant models -- thereby producing a resulting scan that incorporates the proper frequency interval and mode sampling for a given stellar model -- will be presented briefly in this section. 

The properties of a mode is characterised by its order $n = n_{\rm p} - n_{\rm g}$, where $n_{\rm p}$ and $n_{\rm g}$ are the orders associated with the acoustic and gravity-mode cavities, as introduced in Sect. ~\ref{sec:RedGiantOsc}. Depending on the evolutionary state of the star, the relevant range in $n$ progresses to increasingly negative values as $n_{\rm g}$ rises. Defining a suitable interval in $n$, in which to search for modes, must therefore be tied to the evolutionary state of a given star. In order to ensure that this interval encapsulates the entire range of possible observable modes, a conservative lower boundary is set. The lower boundary is defaulted to $n=-2000$ and the upper to $n=50$. The lower boundary is gradually altered according to the steady evolution in $\nu_{\mathrm{max}}$ through a simple step-function, always opting for the conservative choice and ensuring proper sampling. As we approach the TRGB, the lower boundary goes as low as $n=-20 \ 000$, while the upper boundary remains unchanged at all times. It is readily apparent that such a large interval in $n$ for highly evolved red-giant stars results in a plethora of possible theoretical oscillation frequencies for calculation.

In order to resolve such dense oscillation spectra, the number of numerical mesh points, $\mathcal{N}$, must be increased accordingly in order to trace the structure of these evolved stars. Providing a suitable $\mathcal{N}$ aims to ensure that the accuracy of the numerically derived modes remains higher than the typical uncertainty on the observed frequencies of giant stars -- which is of the order of $10^{-4}$ $\mu$Hz (\citealt{Corsaro15}; \citealt{Montellano18}). However, a strict treatment for the required $\mathcal{N}$ as a function of the evolutionary state of a given stellar model has not yet been obtained. Within this work, the number of meshpoints was therefore doubled at specific points along the RGB. Being required to increase the number of meshpoints $\mathcal{N}$ is one of the main components that significantly increases the computation time of the modes for giant stars, in combination with the density of the oscillation spectra themselves.

The settings outlined above do provide suitable giant-star oscillation spectra, but at a considerable computational cost and with a dense mode spectrum associated with each stellar model. Figure~\ref{fig:FullModEchelle} shows an example of such an expensive "full calculation" for model $\mathcal{M}$ in a replicated {\'e}chelle diagram \citep{Bedding2012}. The smear of overlapping oscillations consist of a combination of $l=1,2$ modes that take on mixed-mode characteristics, and visualises the large host of theoretical modes present. Each line of points correspond to a specific radial-mode order $n_{\rm p0}$. Even for such a model below the RGB bump, the calculation time necessary is upwards of $5$ minutes\footnote{Computed on a single CPU allocated on a given node on the Grendel cluster at the Centre for Scientific Computing, Aarhus, \url{http://www.cscaa.dk/grendel/hardware/}}, increasing rapidly as $\Delta\nu$ decreases. For post-RGB-bump models, the dense nature of the modes seen at the lower frequencies in Fig.~\ref{fig:FullModEchelle} will cover the entire frequency range for both the dipole and quadrupole modes. The calculation time necessary for such models reach a couple of hours per model near the TRGB.

\begin{figure}[t]
    \resizebox{\hsize}{!}{\includegraphics{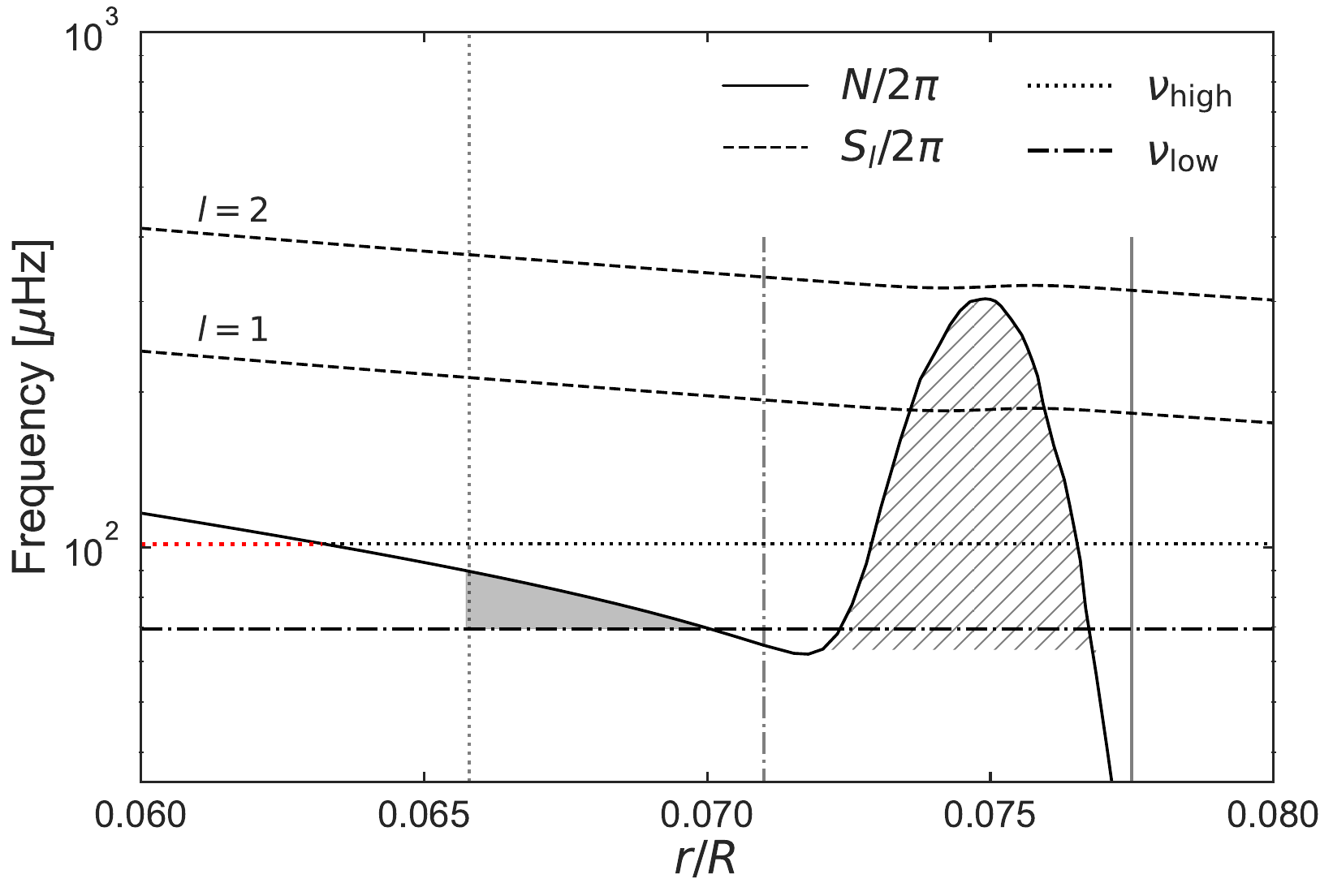}}
    \caption{Propagation diagram for model $\mathcal{M}$, displaying a zoomed-in view of Fig.~\ref{fig:CharFreqs} near the bottom of the convective region. The buoyancy frequency $N/2\pi$ and Lamb frequency $S_l/2\pi$ for $l=1,2$ are shown as the full and dashed ($S_2$ being the higher) profiles, respectively. The horizontal dotted line indicates a representative high frequency $\nu_\mathrm{high}$ of the model, with its g-mode cavity marked in red. The horizontal dot-dashed line shows a representative lower frequency $\nu_\mathrm{low}$. The grey vertical dotted and dot-dashed lines indicate suitable respective truncation points in $r/R$. The grey vertical solid line represents a truncation point at the bottom of the convective region. The buoyancy glitch is marked as the hatched area. }
    \label{fig:PropDiag}
\end{figure}

\section{Isolating p-mode-like frequencies}\label{sec:TruncatingModels}
The most likely oscillation frequencies to be observable are the mixed modes that are the most p-mode-like -- the so-called $(p-m)$ \emph{modes} \citep{Mosser15} -- as the acoustic cavity of the p-modes resides in the exterior of the star where the density $\rho$ is lower, leading to the modes obtaining a lower mode inertia \citep{RGCII}. Theoretically isolating these modes was done through a truncation in the interior of the input model. A truncation is in other words "cutting off" part of the model, making it inaccessible for the pulsation code and thereby removing its possible influence on the oscillation modes. This framework requires an inner truncation point to be given in an evanescent region. If this is satisfied, we can isolate the acoustic resonances as those which exhibit eigenfunctions that are exponentially decreasing towards the interior (\citealt{AsteroACDK}, Chap. 3). This requirement will become satisfied directly by the implemented truncation algorithm and is illustrated in Fig.~\ref{fig:PropDiag}. For details on the truncation algorithm we refer to Appendix \ref{app:TruncAlg}. 

Figure~\ref{fig:PropDiag} displays a propagation diagram for model $\mathcal{M}$, providing a zoomed-in perspective of Fig.~\ref{fig:CharFreqs}. It depicts the interior structure of $N$ and $S_l$ near the bottom of the convective region, where the g-mode cavity has its boundary. The truncation point obtained for a representative frequency just above $\nu_{\mathrm{max}}$ of the model is marked by the vertical dotted line, below which the model could subsequently be truncated. However, when including a tentative lower frequency $\nu_\mathrm{low}$ marked by the horizontal dot-dashed line, a problem arises. The shaded grey area marks a possible g-mode cavity for $\nu_\mathrm{low}$ if the truncation point marked by the vertical dotted line is employed, resulting in the mode obtaining g-mode characteristics. In such cases the implemented frequency-dependent truncation scheme therefore finds a more suitable truncation for $\nu_\mathrm{low}$, indicated by the vertical dot-dashed line, at a higher value of $r/R$. This variable choice of truncation point ensures that we always truncate within an evanescent region and at a sufficiently large separation from $N$ to avoid the influence of g-mode characteristics on a given mode. Subsequently, we are thus able to isolate the acoustic resonances by considering the modes whose eigenfunction decreases exponentially towards the interior. The further details and considerations of Fig.~\ref{fig:PropDiag} are discussed in Sect.~\ref{subsec:Glitch}.

\subsection{Frequency shift by truncation}\label{subsec:FreqShift}
The main complication incurred by the model truncation is a frequency shift of the dipole and quadrupole modes. The shift occurs due to the boundary conditions for the numerical scheme outlined in Sect.~\ref{subsec:ADIPLSNumScheme} being unchanged, despite the extent of the model being altered by the truncation. In other words, the inner boundary conditions do not fully represent the behaviour of the modes in a full model, resulting in a small shift in the retrieved frequencies. When wanting to use the recovered acoustic resonances as centrepoints of the observable regions, this shift must be corrected for. To do this in a coherent way becomes crucial for the success of the method, as otherwise the final frequencies fail to properly estimate the lowest inertia modes of a full model calculation. 

For this purpose, the validation grid to be presented in Sect.~\ref{sec:Validation} was employed (see Appendix \ref{app:GarSettings} for details). For all models across the 9 tracks in the grid, both the full model frequencies calculated with the settings outlined in Sect.~\ref{sec:CalculatingGiantOsc} and the truncated frequencies are at hand. The specifics of deriving the frequency-shift correction can be found in Appendix \ref{app:FreqCorr}. It is concluded that the frequency shift becomes relevant for the dipole modes and has a slight (and predictably consistent) effect on the three lowest acoustic-mode-order quadrupole resonances. Furthermore, it was found that the uncertainty on the corrected resonances to estimate the full model resonances was negligible within the framework of this method, for all but the lowest acoustic-mode-order dipole resonance. Proper recovery of the lowest acoustic dipole resonance is thus problematic, a complication to be discussed and dealt with in Sect.~\ref{sec:ScanningIntervals}.

\subsection{The buoyancy glitch}\label{subsec:Glitch}
In pre-RGB-bump models the phenomenon of a buoyancy glitch occurs. The glitch is a real physical effect left behind by the first dredge-up, producing a discontinuity in composition and hence density in the affected models \citep{CD_15, Lindsay22}. This can cause problems in the application of the Truncated Scanning Method. The variable truncation algorithm initially neglects the glitch and assumes a smooth decrease of $N$ with increasing $r/R$ in the stellar interior. An example of a prominent buoyancy glitch can be seen in the propagation diagram in Fig.~\ref{fig:PropDiag} as the hatched area. This may give rise to a mode essentially corresponding to the interface mode at a density discontinuity \citep{Dziemb_91, Bildsten_98}, with a frequency increasing with degree. This frequency may be in the range of frequencies searched for acoustic resonances. 

In the variable truncation algorithm, modes in the truncated model are tested for being associated with the glitch. A tentative glitch effect on a given mode calculated with the variable truncation algorithm will reveal itself through an oscillating eigenfunction in the glitch region. If a node in the eigenfunction is encountered within the glitch, the given truncated mode is recalculated with the truncation point set at the bottom of the convective region, where the buoyancy frequency from Eq.~\ref{eq:BuoyancyFreq} becomes imaginary. This effectively means placing the truncation point on the exterior side of the glitch, as illustrated by the vertical solid line in Fig.~\ref{fig:PropDiag}, thus eliminating the possible glitch effects. 

Placing the truncation point at the bottom of the convective region may be problematic due to the proximity of $S_l$. Recalling that we must ensure a truncation within an evanescent region, we check for this tentative situation. If the truncation at the bottom of the convective region is invalid due to the evanescent criteria, we reverse to the previous truncation on the interior of the glitch. In such cases, we have to suffer the possible interference with the glitch on the mode. 

We note that the induced effect of the buoyancy glitch on stellar pulsations is still an active research area. This affects not just our truncated models, but also the full pulsation calculations of red-giant models in general. As described by \citet{Cunha24}, the modelling of the glitch and its assumed shape in the stellar interior imprints various features onto the pulsations. It is beyond the scope of this work to investigate these effects further, and we choose to model the glitch as either a $\delta$-function or a resolved Gaussian profile, depending on the resolution of the glitch in the given stellar model.

In summary, the tentative glitch interference on the truncated models affect a minuscule number of models when applying the method to entire stellar grids. Furthermore, the framework to be described in Sect.~\ref{sec:ScanningIntervals} absorbs the interference such that no model in the validation grid of Sect.~\ref{sec:Validation} exhibits an incomplete oscillations spectrum.

\section{Scanning in observable intervals}\label{sec:ScanningIntervals}
While the acoustic resonances have been estimated, the diagnostic potential of the mixed modes now has to be realised and included. The aim is to place appropriate scanning intervals around the recovered acoustic resonances for both the dipole and quadrupole modes, thereby yielding a span of theoretical mixed-mode frequencies for each acoustic mode order $n_{\rm p}$ that is centred on the feasibly observable regions. Crucially, these scanning regions are computed using the full un-truncated model and as such no interference from a truncation is present in the final frequencies we obtain. Thus, the resulting oscillation frequencies are affected by any tentative structural features present, e.g. the buoyancy glitch. The placement of suitable scanning intervals is the focus of this section. 

A complication was described in Sect.~\ref{subsec:FreqShift}, namely the inability of the generalised frequency correction to estimate the acoustic resonance for the lowest-order dipole mode (see Fig.~\ref{fig:GenCorrInertia} and discussion thereof). An attempt to place a scanning interval here risks missing the bottom of the inertia valley, while also being the most computationally expensive dipole interval to calculate. However, a redeeming aspect is that the lowest acoustic resonance that theoretically exists is well outside of the observable frequency range. For example, for model $\mathcal{M}$ the lowest resonance lies near $20$ $\mu$Hz. This is significantly below the observable region for stars exhibiting a $\Delta\nu\approx8$ $\mu$Hz, having a value of $\nu_\mathrm{max}\approx 80$ $\mu$Hz. Within the framework of the Truncated Scanning Method, the lowest-order dipole acoustic resonance has therefore been excluded from calculation.

For models evolved beyond $\Delta\nu \lesssim 5 \ \mu$Hz the condition of numerical precision imposes a limit on the number of nodes in the eigenfunction relative to the maximum number $\mathcal{N_{\rm max}}$ of meshpoints in the computation, and hence a lower limit on the frequency of quadrupolar modes. This clearly does not affect the calculation for the truncated model. Thus the lowest-order quadrupolar acoustic resonance has to be discarded before proceeding to the scanning interval treatment.

We desire to, in a sense, fill out the bottom of the inertia valleys depicted in Fig.~\ref{fig:InertiaPlot} for every acoustic resonance in a given model. For this purpose, we employ Eq.~\ref{eq:ObsWidth} repeatedly for every acoustic resonance recovered from the truncated run -- apart from the two exceptions mentioned above. Eq.~\ref{eq:ObsWidth} is applied twice-over; once with wide intervals ensuring we capture the acoustic resonances of the full model -- then again to sort and trim the intervals based on the recovered resonances.

\begin{figure}[t]
    \resizebox{\hsize}{!}{\includegraphics{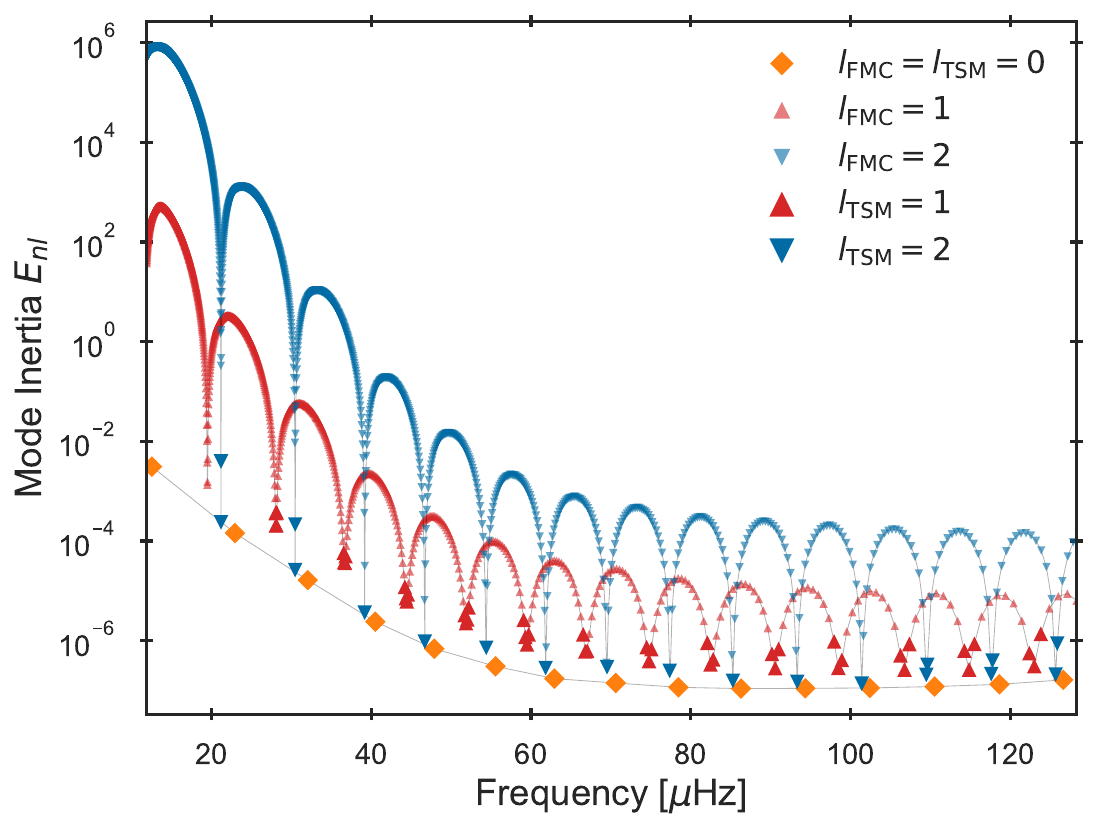}}
    \caption{The mode inertia for the model $\mathcal{M}$, depicting the modes from a full model (transparent points) and Truncated Scanning Method (opaque points) calculation, abbreviated FMC and TSM, respectively. All plotted points are connected by a solid grey line to guide the eye. The inertia valleys containing observable p-mode-like oscillations for both the dipole (red) and quadrupole (blue) modes are recovered. The lowest-order dipole mode has been removed from consideration, as can be clearly seen at $\nu\approx 20$ $\mu$Hz. Note that the recovered radial modes are identical between a FMC and TSM calculation.}
    \label{fig:FinalMethodInertiaPlot}
\end{figure}
\begin{figure*}[t]
    \includegraphics[scale=0.56]{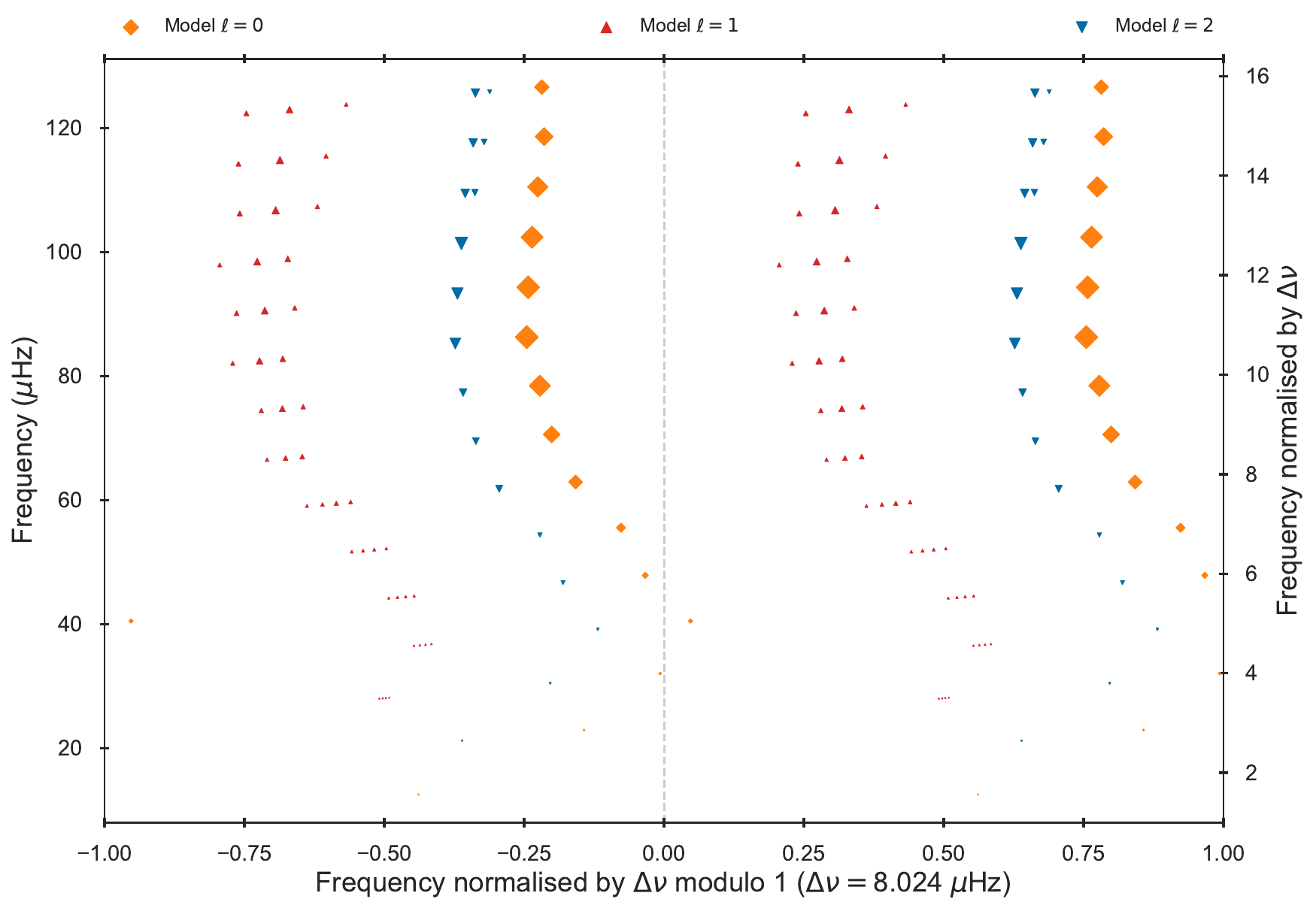}
    \centering
    \caption{Replicated {\'e}chelle diagram of model $\mathcal{M}$ obtained after application of the Truncated Scanning Method for calculating red-giant oscillations. The orange diamonds, red upwards triangles and blue downwards triangles depict the $l=0$, $l=1$ and $l=2$ modes, respectively. A scaling in size related to the inverse of its inertia is applied to each plotted mode. The observable mixed-mode regions surrounding the acoustic resonances have been calculated and plotted for the dipole and quadrupole modes. For contrast to the initial situation compare to Fig.~\ref{fig:FullModEchelle}.}
    \label{fig:TruncModEchelle}
\end{figure*}

\begin{table}[]
\centering
\caption{Classification regions of first-ascent giant-star evolution and their associated amplification factors.}
\begin{tabular}{lcc}
\hline
\multicolumn{1}{c}{\thead{Classification by \\ $\Delta\nu$ of model ($\mu$Hz)}} & \multicolumn{1}{c}{\thead{$l=1$ amplification}} & \multicolumn{1}{c}{\thead{$l=2$ amplification}} \\ \hline 
LRGB: $ \ \ \Delta\nu>8$     &  4              & 5                  \\
MRGB:  $\ \ 8>\Delta\nu>2$   &  4              & 7                  \\
TRGB:  $\ \ \Delta\nu<2$     &  8              & 15                 \\
\hline
\end{tabular}
\label{tab:ScanningIntervals}
\end{table}

\subsection{Initial and trimmed intervals}\label{subssec:InitialIntervals}
Importantly, we cannot risk missing the bottom of each inertia valley in the pulsation calculations. This may occur for e.g. quadrupole frequencies of low mode order where the observable width determined from Eq.~\ref{eq:ObsWidth} becomes so small -- can fall below $1\times 10^{-4} \ \mu$Hz -- that only the resonance itself is observable. The risk of missing the full model resonance, hereafter true resonance, based on the truncated resonances thus becomes significant. To err on the side of caution, the observable region estimated by Eq.~\ref{eq:ObsWidth} is amplified by a gradually increasing scaling factor for the LRGB, MRGB and TRGB for both the dipole and quadrupole cases. The classification of the evolutionary regions and the amplification factors are seen in Table \ref{tab:ScanningIntervals}. The evolutionary region of key importance for future investigations is represented by the MRGB classification, effectively covering an extended region around the RGB bump. It represents the region where our method starts to offer significant computational improvement (see Sect.~\ref{sec:Results}). 

As a final safeguard, we enforce a minimum width of 2 and 1 $\%$ of the $\Delta\nu$ value of the model for dipole and quadrupole modes, respectively. This ensures that we recover the true resonance in the scan, even for complex model interiors where the procedure of Sect.~\ref{subsec:Observability} struggles as well as in situations where the observable width becomes incredibly narrow. We note that the $\Delta\nu$ safeguard is employed often for the quadrupole frequencies of models in the MRGB and TRGB. 

As previously mentioned, the impact of the frequency shift on the quadrupole modes only affect the three lowest acoustic mode order resonances, and always with a consistent small shift to slightly higher frequency. To account for this, we therefore apply twice the width -- either amplified or set by $\Delta\nu$ -- to the lower boundary of the scanning interval of these three modes. 

The calculations are now performed, yielding an initial frequency list for a given stellar model. The recovered intervals are too wide when considering the limitations on the observability discussed in Sect.~\ref{subsec:Observability} and a final assessment is therefore made. Within each obtained interval we locate the resonance as the minima in mode inertia. Employing these true acoustic resonances in Eq.~\ref{eq:ObsWidth} provides the \emph{actual observational regions of the complete stellar model}. Due to the asymmetry of the frequency spacing -- i.e two modes/resonances of comparable inertia at the bottom of an inertia valley, where the minimisation routine chooses one -- as well as the applicational uncertainties and a tentative metallicity dependence of Eq.~\ref{eq:ObsWidth}, we amplify the observational width obtained from Eq.~\ref{eq:ObsWidth} by a factor of 2. The initial wide intervals are subsequently trimmed to these new widths.

Figure~\ref{fig:FinalMethodInertiaPlot} shows the obtained result for model $\mathcal{M}$ employing the above procedure in a plot of the mode inertia. It is clear that the bottom of inertia valleys are recovered by the scanning intervals. Furthermore, we obtain a varying number of mixed modes of incrementally higher inertia on each side of the acoustic resonances at the minima in accordance with the observational limitations. In some mode orders, only the single quadrupole resonance is recovered. At face value it seems excessive to have conducted an initial scanning for a wide interval, but the determination of even this single resonance was previously unattainable.

\subsection{LRGB stars}\label{subsec:LowEvo}
For stars exhibiting a $\Delta\nu \gtrsim 8$ $\mu$Hz, the mode coupling $q$ can become very strong and reach the upper limit of $1/4$. Additionally, as the radiative damping is as yet inefficient, this results in mixed modes with a considerable shift in frequency becoming observable. At higher mode orders, this effectively means that the amplified observational width determined for the initial scanning intervals crosses over the inertia bump and overlaps. This is unproblematic, as for the LRGB members, the number of modes at the inertia peaks at higher frequency is modest. Encountering this situation the intervals are simply joined into a complete scan covering the relevant region in frequency. A visual example akin to Fig.~\ref{fig:FinalMethodInertiaPlot} can be seen in Fig.~\ref{fig:UnevolvedStarInertiaPlot} for a LRGB star with $\Delta\nu=14.86$ $\mu$Hz, illustrating the reduced number of modes at higher acoustic mode order at the inertia peaks, and how we recover significantly displaced observable mixed modes.

\subsection{TRGB stars}\label{subsec:HighEvo}
Highly evolved red giants also deserve special mention. Here, the coupling strength has become very low in combination with efficient radiative damping, such that only the true resonances or mixed modes very close to it are observable. However, due to the nature of red-giant oscillations outlined in Sect.~\ref{sec:RedGiantOsc}, an immense number of theoretical modes exist. Figure~\ref{fig:HighlyEvoStarInertiaPlot} shows an example of a TRGB star with $\Delta\nu=1.04$ $\mu$Hz. Despite the observational limitations, we still recover a small number of mixed modes around the true resonances. Yet, they are displaced very little in frequency and would likely overlap in observations to provide a widened frequency peak around the acoustic resonance, as found by \citet{Stello14}.

\section{Validation}\label{sec:Validation}
In order to validate the performance of the Truncated Scanning Method, it was applied to a set of stellar tracks covering a large parameter space. A cartesian grid of 9 tracks was calculated with masses $M=$ 0.7, 1.3 and 1.9 M$_\odot$ and metallicities $[\mathrm{Fe}/\mathrm{H}]=$ $-1.5$, $-0.5$ and $0.5$ dex with GARSTEC \citep{Achim08}. For brevity, the details of the input to and settings within GARSTEC can be found in Appendix \ref{app:GarSettings}. For each of the tracks, stellar models were recorded from $\Delta\nu\approx 20 \ \mu$Hz until the Helium flash at $\Delta\nu\approx 0.8 \ \mu$Hz, providing an appropriate sampling for testing purposes of $\sim100$ models per track along this evolutionary region. 

The complete Truncated Scanning Method framework described in Sects.~\ref{sec:CalculatingGiantOsc} - \ref{sec:ScanningIntervals} was applied to all models in the grid. Additionally, the computationally expensive full model calculations were also obtained for the entire grid as to provide grounds for proper validation of the Truncated Scanning Method. An equivalent to Fig.~\ref{fig:FinalMethodInertiaPlot} was inspected for every model, enabling visual assessment and verification of the performance of the method. 

The Truncated Scanning Method performed as intended for models across the large parameter space covered by the validation grid. For a limited number of models, the glitch interference causes the obtained observational regions to take peculiar shapes -- as briefly mentioned in \ref{subsec:Glitch} -- but nonetheless localise the observable regions with the acoustic resonances. Generally, the wide applicability of our method is verified for solar-like stars on the RGB and confidently provides an avenue for obtaining the oscillations of red-giant models.

\section{Results}\label{sec:Results}
Employing the Truncated Scanning Method we can compute the realistically observable theoretical mixed-mode oscillations of red-giant models at significantly reduced computational cost. The result for model $\mathcal{M}$ is displayed in a replicated {\'e}chelle diagram in Fig.~\ref{fig:TruncModEchelle}, akin to the full model version shown previously in Fig.~\ref{fig:FullModEchelle}. Figure \ref{fig:TruncModEchelle} shows the resulting scanning intervals in a more intuitive way in connection to observational asteroseismology. The observable regions determined return the modes adjacent to the acoustic resonance of each mode order, distributing the modes along the approximately vertical ridges usually found in {\'e}chelle diagrams of observed pulsations. We see how the determined dipole mixed modes reside in the region with the lowest inertia for the model, as indicated by the size of the plotted mode markers. For model $\mathcal{M}$, the observable quadrupole frequencies number just the acoustic resonance in each mode order -- except for the highest frequencies. These singular quadrupole resonances were previously unattainable and hidden among the multitude of theoretical frequencies, but are now recovered by the Truncated Scanning Method. The resulting theoretical mode spectrum obtained for a given model is thus situated as desired in the regions crucial for investigations that fit observed stellar data. 

It is important to acknowledge that due to the complications with the variable truncation scheme and glitch interference outlined in Sect.~\ref{sec:TruncatingModels}, a small number of stellar models -- primarily in the LRGB region -- will inevitably fail in their computation. Such models must be recalculated in their full form with the settings outlined in Sect.~\ref{sec:CalculatingGiantOsc}. However, for the few affected and less evolved models, the computation remains sufficiently efficient to perform.

\begin{figure}[t]
    \resizebox{\hsize}{!}{\includegraphics{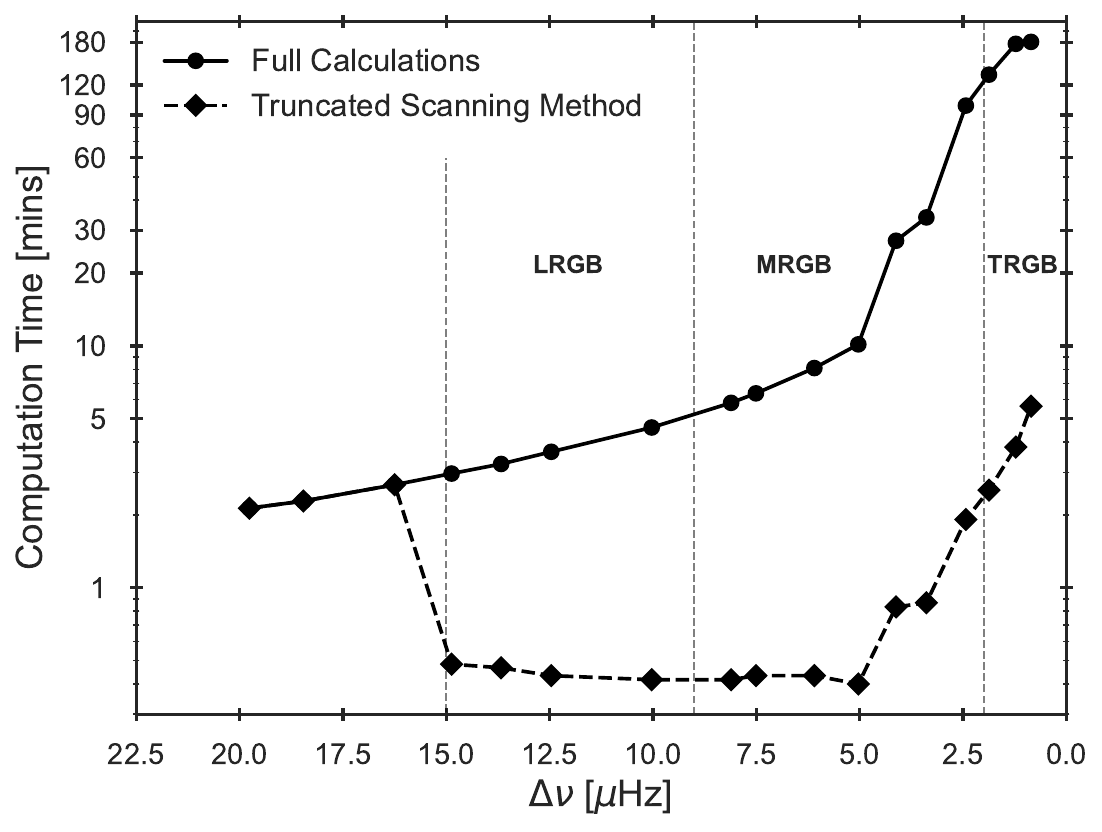}}
    \caption{Computational efficiency of the Truncated Scanning Method. The computation time for a given model is plotted against its $\Delta\nu$ value for a full calculation and for the Truncated Scanning Method as the solid and dashed lines, respectively. Note that the time is displayed on a logarithmic scale and the overplotted lines are simply for aiding in visualisation. The vertical dashed lines indicate the classification transition points from Table \ref{tab:ScanningIntervals} as shown.}
    \label{fig:CompEff}
\end{figure}

Prior to this work, the only option at hand was to calculate the entire frequency spectra of evolved stars. As became clear in previous discussions, these incredibly dense spectra -- and hence large number of nodes in the mode eigenfunctions requiring a significant increase in the number $\mathcal{N}$ of numerical meshpoints  -- results in excessive computational costs. Briefly outlining the obtained computational efficiency puts into perspective the importance and applicability of the Truncated Scanning Method. The full model calculation in Fig.~\ref{fig:FullModEchelle} takes approximately 5 minutes to calculate, while the results obtained from the Truncated Scanning Method in Fig.~\ref{fig:TruncModEchelle} was computed in 16 seconds. This computational difference only becomes increasingly more significant during the ascent of the RGB. For highly evolved red-giant stars near the TRGB, a full model calculation of a single stellar model takes hours, while the Truncated Scanning method can perform the calculation in under $10$ minutes per model.

Figure~\ref{fig:CompEff} illustrates the obtained efficiency for a number of models from the solar origin track of model $\mathcal{M}$ selected in each of the classification intervals from Table \ref{tab:ScanningIntervals}. Prior to the method being applied at $\Delta\nu=15 \ \mu$Hz the computation time is identical, but varies significantly once the Truncated Scanning Method is employed. The calculation time for the Truncated Scanning Method varies slightly along the LRGB and MRGB due to the dynamic interval boundaries that depend on the specific model, from the considerations of Sects.~\ref{subsec:Observability} and \ref{sec:ScanningIntervals}. 

Until we reach the TRGB at $\Delta\nu<2 \ \mu$Hz, the calculation time with the Truncated Scanning Method remains below $\sim2$ minutes, which is an impressive improvement to the previous situation. Furthermore, it speaks to the applicability and performance of the derived method. For the TRGB models, the computation time exceeds $2-3$ hours for the full calculations, but at all times remains below $\sim8$ minutes for the Truncated Scanning Method. The appearance of a reduced slope for the full calculations near the TRGB is due to the logarithmic scale of the y-axis. It does not reflect an actual difference in the scaling of the computation time between results produced by a full calculation and the Truncated Scanning Method.

\section{Discussion and conclusion}\label{sec:DiscussionAndConclusion}
The Truncated Scanning Method contributes a novel approach to deriving the oscillation spectra of evolved giants efficiently, by limiting the derivation to the observable regions. However, as with any scientific method in its infancy, there is room for future improvement to the method outlined herein. 

The simplicity of the method is reduced by the need for a frequency-shift correction of the dipole modes. The shift in frequency impacted all recovered acoustic resonances since the truncation of the model and the applied inner boundary condition did not fully reflect the properties of modes in the full model. While this frequency shift is small in magnitude, avoiding it would be preferable. In seeking out a way to avoid the frequency shift, the entire procedure could be redefined to avoid the necessity of a truncation altogether. Through reformulation of the equations behind the stellar oscillations, one could discard the contributions that lead to g-mode behaviour following the ideas proposed in \citet{Ong20}. Further studies are required, but it may lead to increased accuracy and simplicity by removing the need for a truncation and the accompanied frequency shift from the Truncated Scanning Method. The potential of this approach is the subject of further testing and development efforts. 

The induced signatures on the pulsations caused by the buoyancy glitch cause problems for a very limited number of models in the evolutionary region where the glitch is dominant (when the convective region is receding). We stress that the effect of the glitch is not unique to the approach in the Truncated Scanning Method, but is a general effect imposed on all oscillation calculations of the affected red-giant models. Constraining the glitch signatures and understanding their nature is an important avenue for further research and improvement for red-giant pulsations. 

The results of this paper serve as a novel method for efficiently calculating stellar oscillations for stellar models of giant stars. The desired results of a much faster computation that maintains reliability in the returned theoretical frequencies spaced across the observational regions for any given red-giant model was obtained. Realising these computations unlocks the possibility of asteroseismic modelling utilising individual frequencies of red-giant stars, which could serve as an independent modelling procedure to the existing reliance on the global parameters. 

On a larger scale, the Truncated Scanning Method moves investigations of individual frequencies by fitting to stellar grids situated on the RGB into the realm of possibility. The calculation time of all pulsation modes for an entire grid of evolved stellar tracks was previously infeasible both in terms of computational resources and time invested for a certain research project, but has now become achievable in a matter of days as opposed to months or even years.  In this context, the utilisation of the Truncated Scanning Method for an entire stellar grid in connection to asteroseismic modelling of red giants with individual frequencies is the subject of an upcoming paper.

\vspace{-2mm}
\section*{Data and software products}
The code products and required software that produces ADIPLS input files with the Truncated Scanning Method settings are available upon request to the first author. Various versions of the validation grid with the stellar models and associated frequency files used for the representative figures, along with the plotting code, can be found here: \url{https://www.erda.au.dk/archives/a6d160238869305894269d9a74068333/published-archive.html}.

\begin{acknowledgements}
    JRL wishes to thank the members of SAC for comments and discussions regarding the paper. The authors thank the anonymous referees for the constructive criticism, which aided in improving the quality of the work in this paper. This work was supported by a research grant (42101) from VILLUM FONDEN. Funding for the Stellar Astrophysics Centre was provided by The Danish National Research Foundation (grant agreement no.: DNRF106). MSL acknowledges support from The Independent Research Fund Denmark's Inge Lehmann  program (grant  agreement  no.:  1131-00014B). The numerical results presented in this work were obtained at the Centre for Scientific Computing, Aarhus \url{https://phys.au.dk/forskning/faciliteter/cscaa/}.
\end{acknowledgements}

\bibliographystyle{aa.bst} 
\bibliography{bibliography.bib} 

\begin{appendix}
\section{Frequency-dependent truncation algorithm}\label{app:TruncAlg}
The choice of truncation point is determined through the following equation from JWKB theory:
\begin{equation}\label{eq:VarTrunc}
    k^2 = \frac{l(l+1)}{x^2}\left(1-\frac{\Tilde{N}^2}{\sigma^2}\right)\left(1-\frac{\sigma^2}{\Tilde{S}_{l}^2}\right)
\end{equation}
Here, $\sigma^2=\frac{R^3}{GM}\omega^2$ is the squared dimensionless frequency with $\omega=2\pi\nu$ describing any angular oscillation frequency of a given mode and $x=r/R$ the radial coordinate. The dimensionless characteristic frequencies are denoted as $\Tilde{N}$ and $\Tilde{S}_l$. The modes have clear acoustic behaviour through the Lamb frequency $\Tilde{S}_l$, i.e. p-mode behaviour, as long as the term $\Tilde{N}/\sigma^2 \ll 1$. However, as $\Tilde{N}$ grows large for giant stars $\Tilde{N}/\sigma^2 \simeq 1$ is reached, resulting in gravity-dominated mixed modes -- so-called $(p-g)$ modes \citep{Mosser15} -- arising, all exhibiting higher mode inertia. If we wish to isolate the observable p-mode-like oscillations, we therefore need to truncate the model according to the behaviour of $\Tilde{N}/\sigma^2$. Note that this fraction is frequency dependent. Hence, the approach is to implement a variable truncation depending on frequency. 

Instead of directly using $\Tilde{N}/\sigma^2$ to define the truncation, the term relating to the Lamb frequency $\sigma^2/\Tilde{S}_{l}^2$ was also considered to provide an upper boundary condition. The aim is to find a truncation point $x_\mathrm{tr}$ as the point where both $\Tilde{N}^2/\sigma^2$ and $\sigma^2/\Tilde{S}_{l}^2$ are well below $1$. A balance between minimising $x$ in relation to
\begin{equation}
    \frac{\Tilde{N}^2(x_\mathrm{tr}^{(1)})}{\sigma^2} \leq \epsilon \ ,
\end{equation}
in order to obtain $x_\mathrm{tr}^{(1)}$, and concurrently maximising $x$ in relation to
\begin{equation}
    \frac{\sigma^2}{\Tilde{S}_{l}^2(x_\mathrm{tr}^{(2)})} \geq \zeta \ ,
\end{equation}
to estimate $x_\mathrm{tr}^{(2)}$ is performed. Currently, we hard-coded the values of $\epsilon=\zeta=0.5$. In the case where $x^{(1)}_\mathrm{tr}\leq x^{(2)}_\mathrm{tr}$, the truncation point can directly be set to $x_\mathrm{tr}=x^{(1)}_\mathrm{tr}$ and a successful truncation is obtained ensuring that
\begin{equation}
    \frac{\Tilde{N}^2}{\sigma^2} \leq \epsilon \ , \ \ \     \frac{\sigma^2}{\Tilde{S}_{l}^2} \geq \zeta \ .
\end{equation}
In the case where the evanescent zone is narrow, the above conditions may not be satisfied resulting in $x^{(1)}_\mathrm{tr} \geq x^{(2)}_\mathrm{tr}$. A last-ditch effort to set $x_\mathrm{tr}=x^{(3)}_\mathrm{tr}$ is evaluated by requiring
\begin{equation}
    \frac{\Tilde{N}^2(x_\mathrm{tr}^{(3)})}{\sigma^2} = \frac{\sigma^2}{\Tilde{S}_{l}^2(x_\mathrm{tr}^{(3)})} .
\end{equation}
For the dipole modes where $N$ and $S_l$ are in closer proximity, this criterion is often employed and sets $x_\mathrm{tr}=x^{(3)}_\mathrm{tr}$. However, if this does not provide a realistic estimate, as there is no guarantee that $k^2>0$ at $x^{(3)}_\mathrm{tr}$ in Eq.~\ref{eq:VarTrunc}, the scan is stopped and an error message to the standard error is produced.

\section{Generalised frequency correction}\label{app:FreqCorr}
\begin{figure}
    \resizebox{\hsize}{!}{\includegraphics{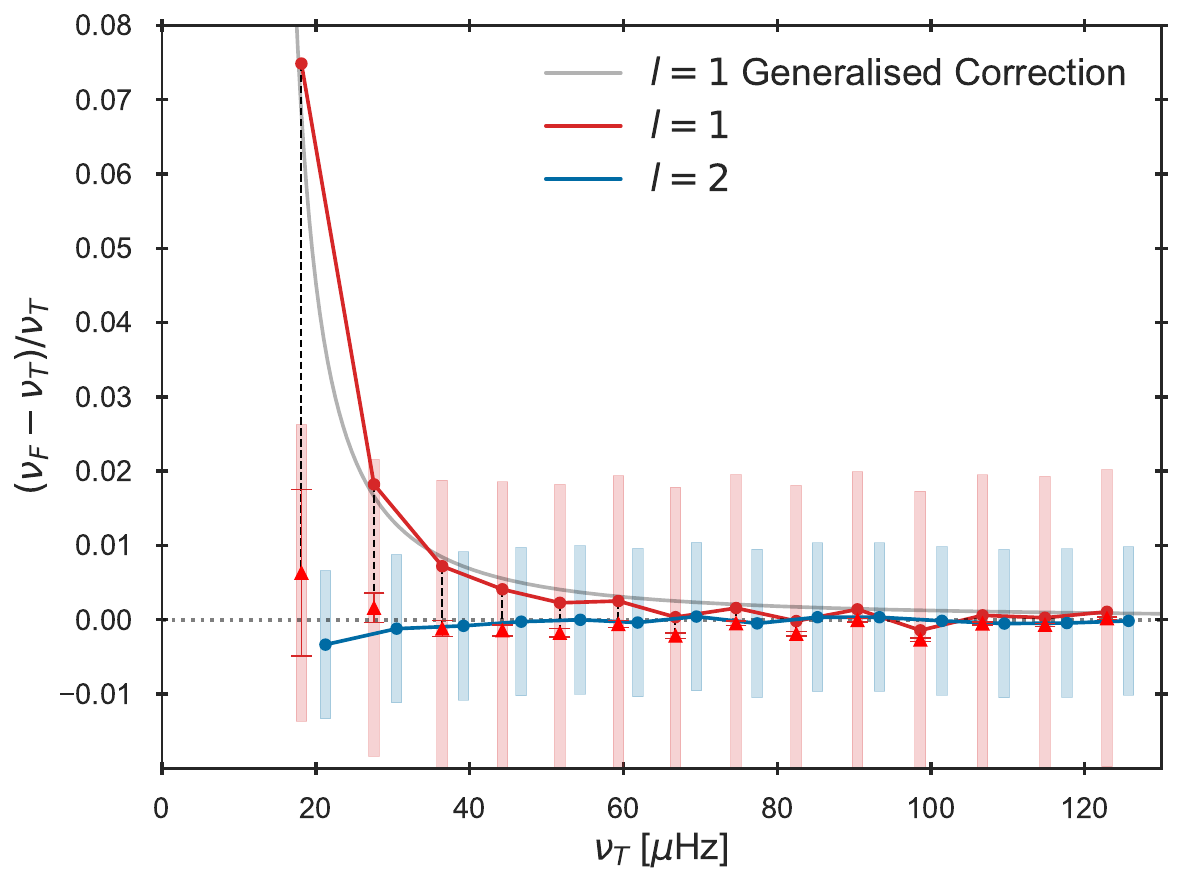}}
    \centering
    \caption{Visual representation of the frequency shift incurred on the truncated frequencies for model $\mathcal{M}$. The fractional frequency shift $(\nu_F-\nu_T)/\nu_T$ between the full ($\nu_F$) and truncated ($\nu_T$) frequencies is plotted against $\nu_T$ for the dipole and quadrupole modes, shown by the red and blue lines, respectively. The grey profile shows the resulting generalised correction for model $\mathcal{M}$ overplotted, for which its value is subsequently subtracted from all dipole acoustic resonances to provide the red triangles. The uncertainty on the correction for each dipole mode is indicated, as well as the minimal size of the scanning interval boundary as the shaded beams in the respective colour.}
    \label{fig:FreqShiftRep}
\end{figure}

Here we consider the diagnostic consequences of the incurred frequency shift mentioned in Sect.~\ref{subsec:FreqShift}, and subsequently the derivation of a generalised model for its correction. For the purpose of the derivation and subsequent testing, the validation grid described in Sect.~\ref{sec:Validation} and Appendix \ref{app:GarSettings} was employed. 

Figure~\ref{fig:FreqShiftRep} illustrates the situation in a relative parameter space for model $\mathcal{M}$, as the fractional difference between the truncated frequencies $\nu_T$ and full frequencies $\nu_F$ recovered at minimum inertia for each acoustic resonance as a function of the truncated frequencies $\nu_T$. One immediately notices a key difference between the effect on the $l=1$ and $l=2$ modes. The frequency shift is almost entirely insignificant for the quadrupole modes, while the contrary is true for certain dipole modes. This is due to the nature of the modes in the stellar interior. The dipole modes have an inner turning point for their p-mode cavity deeper in the interior than the quadrupole modes (see Fig.~\ref{fig:CharFreqs} or \ref{fig:PropDiag}). As such, the $l=1$ modes in a sense `feel´ the effect of the truncation to a larger extent than do those with $l=2$. The necessary frequency shift for the dipole modes, as depicted in Fig.~\ref{fig:FreqShiftRep}, is dominant for the lower-order modes and significantly smaller for the higher-order modes. 

The necessary correction for the quadrupole modes takes a much simpler form. The effect of the small shift seen will generally be absorbed by the inclusion of a frequency scanning interval (see Sect.~\ref{sec:ScanningIntervals}). Only for the three lowest acoustic-mode-order resonances is the shift considerable enough to warrant further consideration. However, the shift is consistent and predictable, resulting in a small shift to larger frequencies. This will be accounted for by the treatment in Sect.~\ref{sec:ScanningIntervals}. The approach presented in the following therefore focuses entirely on the dipole modes of the models.

For all the models in the validation grid it is possible to do an exact fit to the dipole shift trend. However, this is only possible since we have obtained the expensive full model frequencies. The aim is therefore to derive a fitting function that could be generalised and applied widely when the full model calculations are omitted. The fitting function decided upon is of the form
\begin{equation}\label{eq:fitmod}
    f=\frac{1}{(\nu-a)^b} , \ \ \ \  a,b \in \mathbb{R} .
\end{equation}
The approach is to fit Eq.~\ref{eq:fitmod} to the dipole shift of all models in the validation grid, and subsequently investigate how each of the fit parameters varies with the value of $\Delta\nu$ for the given model. All fitting efforts at this stage utilise a non-linear Neyman least-squares fitting algorithm \citep{Neyman49}.

\begin{figure}[h]
    \resizebox{\hsize}{!}{\includegraphics{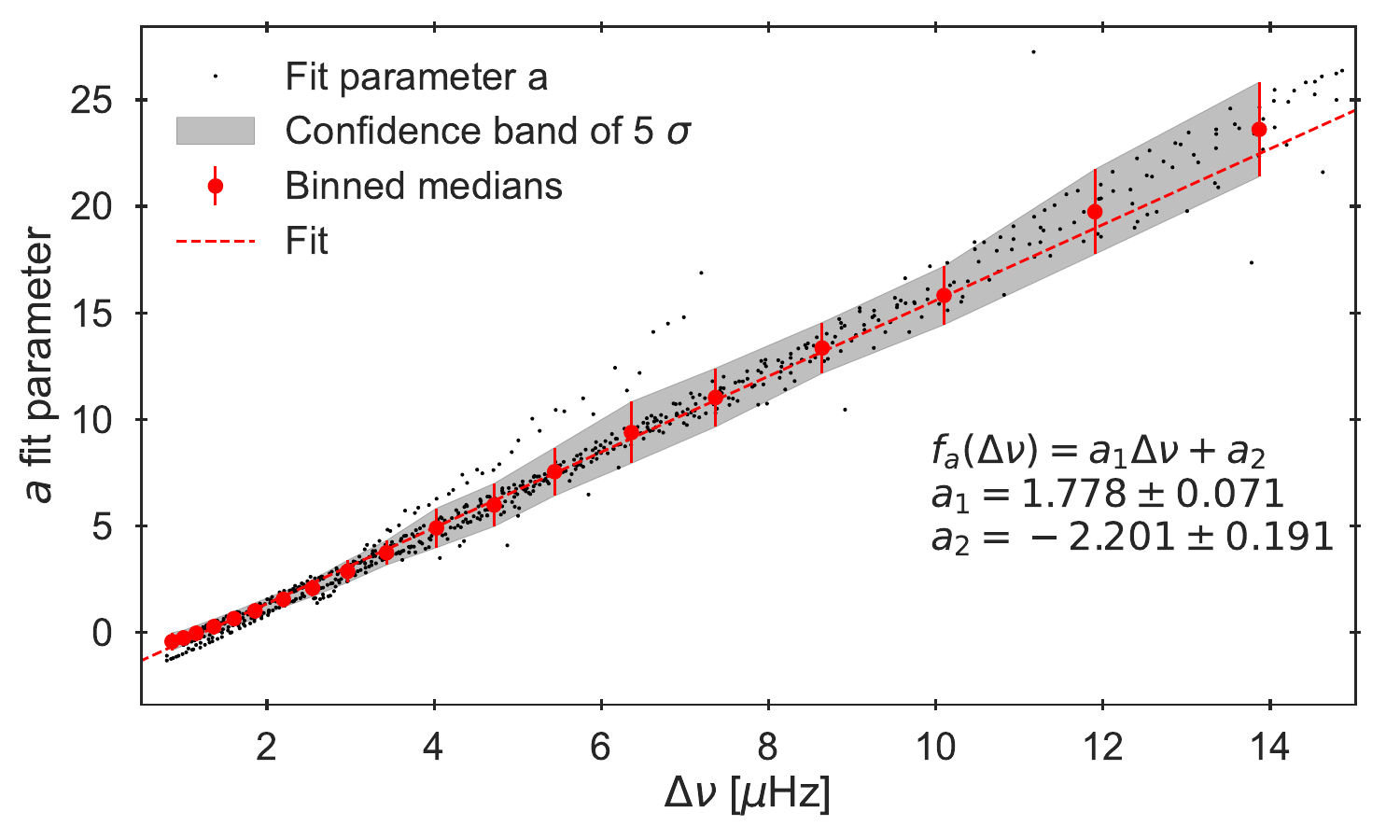}}
    \caption{The fit parameter $a$ obtained from all fits to Eq.~\ref{eq:fitmod} for all models investigated, plotted against $\Delta\nu$. The resulting linear fit to the binned data has been overlaid, with the resulting parameters $a_1$ and $a_2$ indicated.}
    \label{fig:CorrFitaDnu}
\end{figure}
\begin{figure}[h]
    \resizebox{\hsize}{!}{\includegraphics{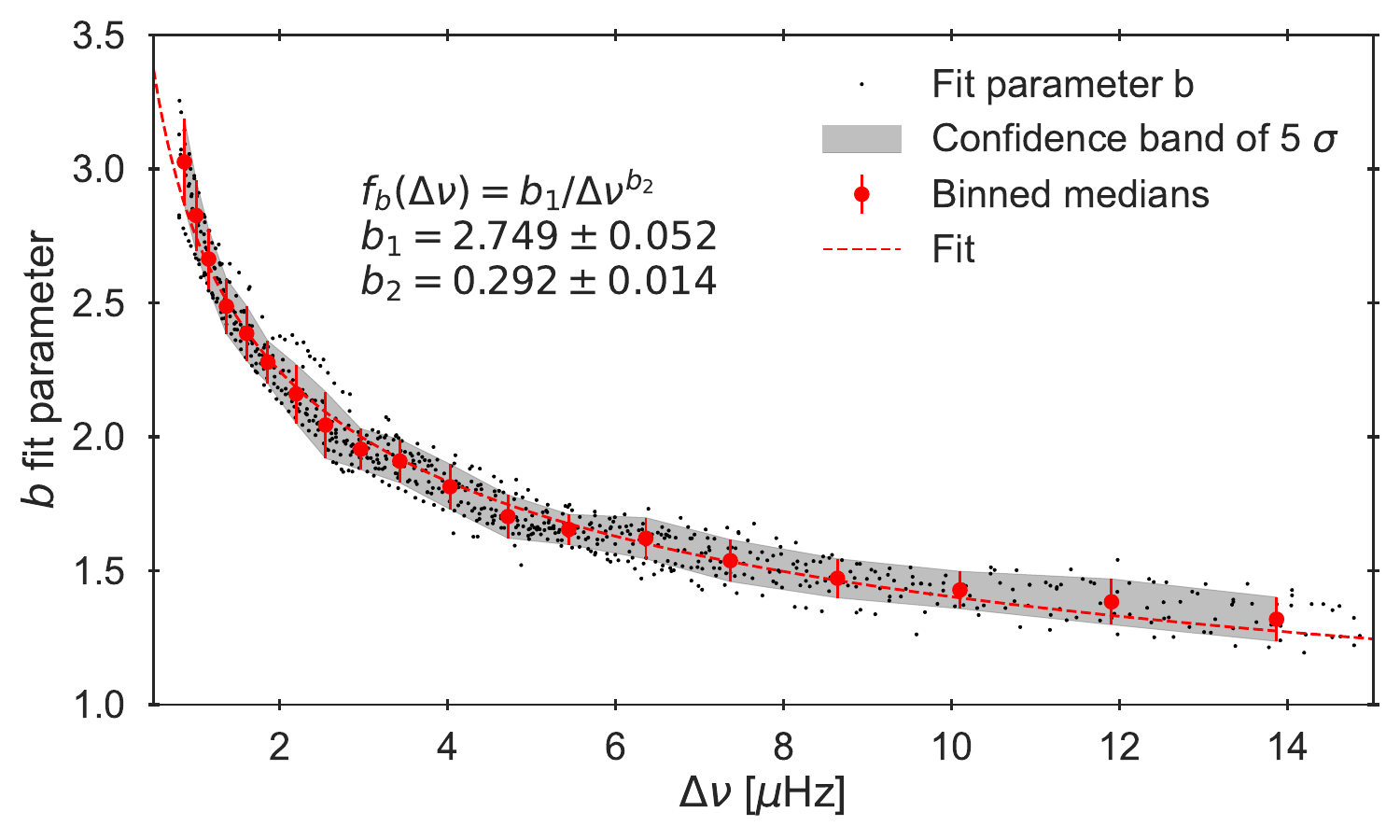}}
    \caption{The fit parameter $b$ obtained from all fits to Eq.~\ref{eq:fitmod} for all models investigated, plotted against $\Delta\nu$. The fit to the binned data using a power function with the indicated parameters $b_1$ and $b_2$ has been overplotted, showing the determined trend of $b$ with $\Delta\nu$.}
    \label{fig:CorrFitbDnu}
\end{figure}
\begin{figure*}[t]
    \includegraphics[width=17cm]{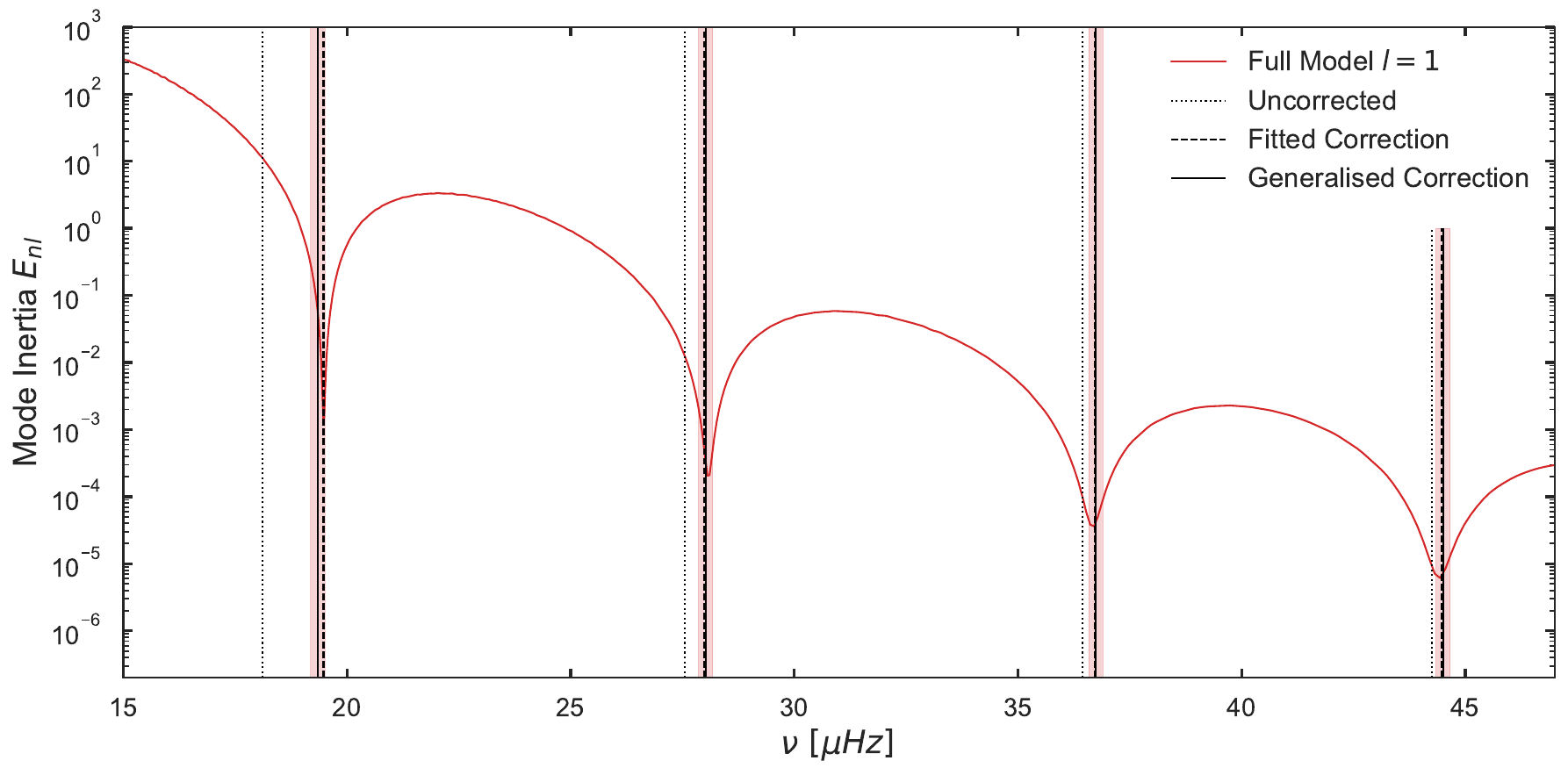}
    \centering
    \caption{Application of the generalised frequency-shift correction for the lower frequency dipole modes of model $\mathcal{M}$ visualised in an inertia diagram. The frequencies of the uncorrected resonance frequencies (dotted lines), corrected resonance frequencies from an exact fit to model $\mathcal{M}$ (dashed lines) and the generalised model-corrected resonance frequencies (solid lines) are shown. Note that the exact and generalised corrections almost overlap in all acoustic resonances. The minimal size of the subsequently imposed scanning interval is indicated by the surrounding shaded bands.}
    \label{fig:GenCorrInertia}
\end{figure*}

The procedure for deriving the general correction based on each model in the validation grid is as follows: 
\begin{enumerate}
    \item Fit the complete fitting function in Eq.~\ref{eq:fitmod} to every model, each with an associated $\Delta\nu$ value. This obtains a range of values for the fit parameters $a$ and $b$ as a function of $\Delta\nu$.
    \item Determine the relation between $a$ and $\Delta\nu$ (Fig.~\ref{fig:CorrFitaDnu}), chosen as a linear fit of the form $f_a=a_1 \Delta\nu + a_2$.
    \item Determine the relation between $b$ and $\Delta\nu$ (Fig.~\ref{fig:CorrFitbDnu}), fitted according to $f_b=b_1/(\Delta\nu)^{b_2}$. 
\end{enumerate}
For points 2 and 3, we bin the respective fit parameters in bins equally distributed in log-space. The median value of the fit parameters within each bin is calculated and the $5-\sigma$ confidence interval used as a measure of the uncertainty. Subsequently we fit the associated function using the \emph{iminuit} package\footnote{\url{https://scikit-hep.org/iminuit/index.html}} with the MINUIT algorithm \citep{MinuitAlg}, which is a fitting tool for rigorous statistical analyses that incorporates the covariances in its uncertainty estimates.  

The resulting generalised model for the frequency-shift correction is then defined as
\begin{equation}\label{eq:GeneralisedCorr}
    f_{\mathrm{corr}}(\nu) = \frac{1}{\left(\nu - f_a\left(\Delta\nu\right)\right)^{f_b(\Delta\nu)}} ,
\end{equation}
where the value of $\Delta\nu$ is in $\mu$Hz. The two correlation functions $f_a$ and $f_b$ are given by the fits in Figures~\ref{fig:CorrFitaDnu} and \ref{fig:CorrFitbDnu}:
\begin{equation}
    \begin{split}
        f_a(\Delta\nu)&=\left(1.778\pm0.071\right)\Delta\nu - \left(2.201\pm0.191\right) \\
        f_b(\Delta\nu)&=\frac{\left(2.749\pm0.052\right)}{\Delta\nu^{\left(0.292\pm0.014\right)}} .
    \end{split}
\end{equation}
Note that for $f_b(\Delta\nu)$, the fitted function displays only a slightly asymptotic behaviour as $\Delta\nu$ approaches zero. This is desirable, as the risk of $b$ being set to unreasonably high values for highly evolved stars is avoided. The uncertainties on the parameters in $f_a$ and $f_b$ are propagated when employing Eq.~\ref{eq:GeneralisedCorr} for model $\mathcal{M}$ and overplotted on the corrected dipole frequencies in Fig.~\ref{fig:FreqShiftRep}. The uncertainties on the corrected frequencies are negligible in comparison to the later employed minimum boundary for the scanning intervals (see Sect.~\ref{sec:ScanningIntervals}), for all but the lowest acoustic-mode-order resonance. For this mode, the correction takes its largest amplitude and is the most uncertain. This complication was also discussed Sect.~\ref{sec:ScanningIntervals}. 

To summarise, we note that the application of the derived generalised frequency correction does not require the full model frequencies, just the recovered acoustic resonances and the $\Delta\nu$ of the model. As such, it can be applied to the dipole modes after a given truncated calculation to correct and shift the recovered acoustic resonances. Figure ~\ref{fig:GenCorrInertia} shows the generalised correction applied to the lower frequency range of model $\mathcal{M}$, visualised through a plot of the mode inertia. One can clearly see how the acoustic resonances have been shifted to be coincident with the inertia valleys as desired -- a larger correction being necessary at lower frequency -- by both an exact fit to model $\mathcal{M}$ and the generalised correction. The previously noted difficulty of the lowest acoustic-mode-order resonance can also be seen in Fig.~\ref{fig:GenCorrInertia}. The minimum interval for the subsequent scanning is indicated by the shaded bands, and shows how the minima in inertia are widely covered for all but this one problematic resonance.

\section{Scanning intervals for LRGB and TRGB stars}
\begin{figure}[h!]
    \resizebox{\hsize}{!}{\includegraphics{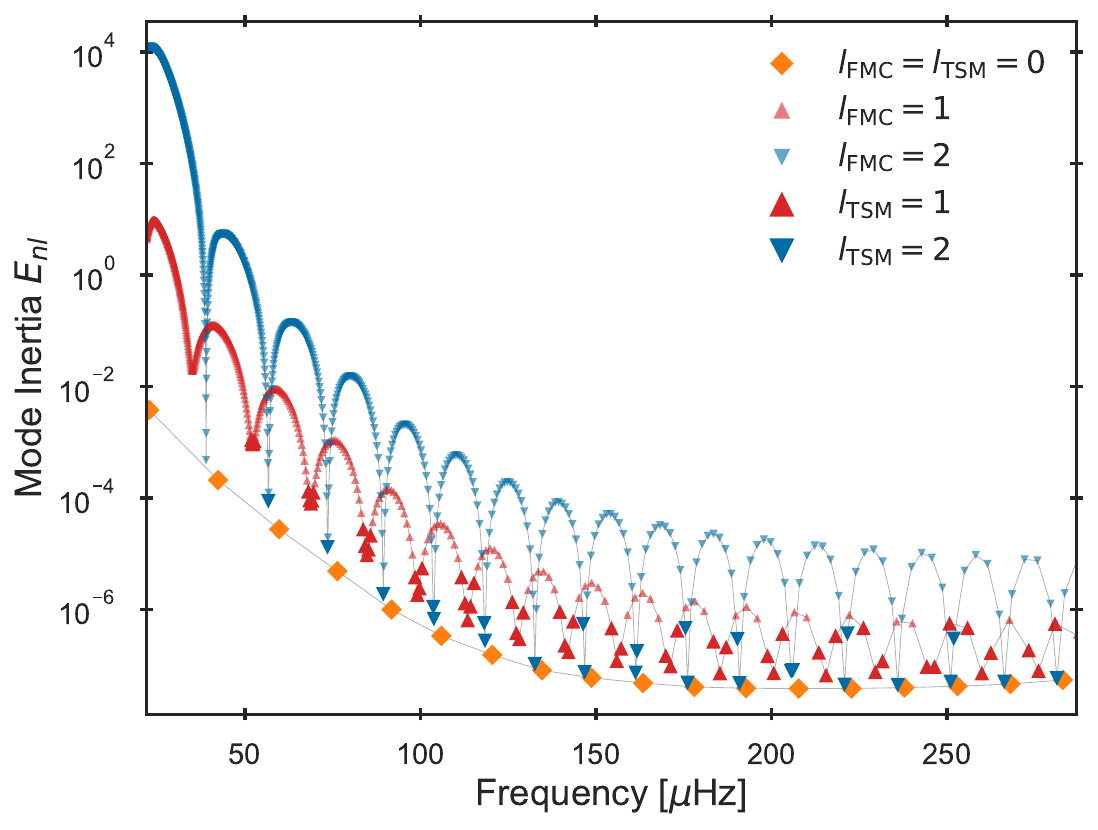}}
    \centering
    \caption{The mode inertia for an unevolved RGB star with $\Delta\nu=14.86$ $\mu$Hz, depicting the modes from a full model (transparent points) and Truncated Scanning Method (opaque points) calculation, abbreviated FMC and TSM, respectively. The observable intervals are recovered for both the dipole (red) and quadrupole (blue) oscillation modes.}
    \label{fig:UnevolvedStarInertiaPlot}
\end{figure}

\begin{figure}[h!]
    \resizebox{\hsize}{!}{\includegraphics{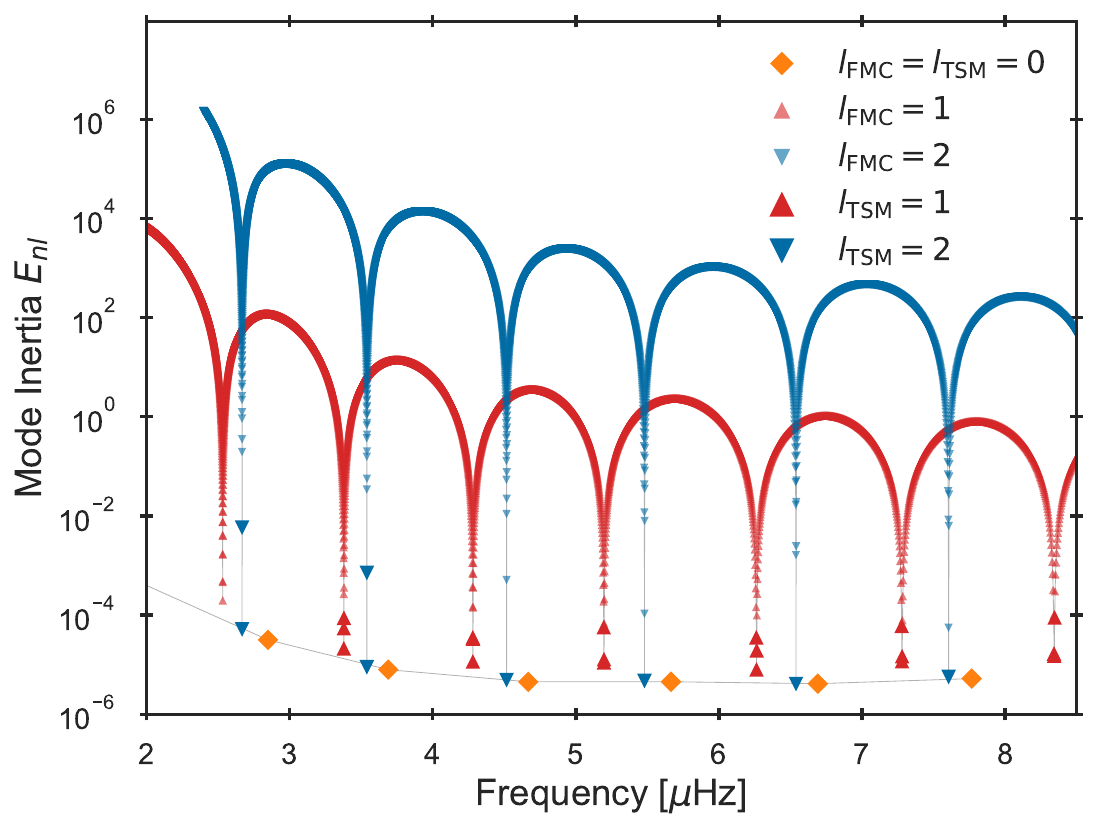}}
    \centering
    \caption{The mode inertia for a highly evolved star with $\Delta\nu=1.04$ $\mu$Hz, depicting the modes from a full model (transparent points) and Truncated Scanning Method (opaque points) calculation, abbreviated FMC and TSM, respectively. The observable regions for both the dipole (red) and quadrupole (blue) modes are recovered, depicting very few observable modes. }
    \label{fig:HighlyEvoStarInertiaPlot}
\end{figure}

Figure~\ref{fig:UnevolvedStarInertiaPlot} and \ref{fig:HighlyEvoStarInertiaPlot} display the mode inertia for an LRGB and TRGB model, respectively. They show the results of the Truncated Scanning Method utilising the altered dimensions of the scanning intervals for the two classification regions mentioned and discussed in Sect.~\ref{subsec:LowEvo} and \ref{subsec:HighEvo}, respectively. The differences of the scanning interval dimensions and resulting observable regions for the LRGB and TRGB evolutionary regions, is visualised in the figures. 

\section{GARSTEC settings for the validation grid}\label{app:GarSettings}
\begin{figure}[]
    \resizebox{\hsize}{!}{\includegraphics{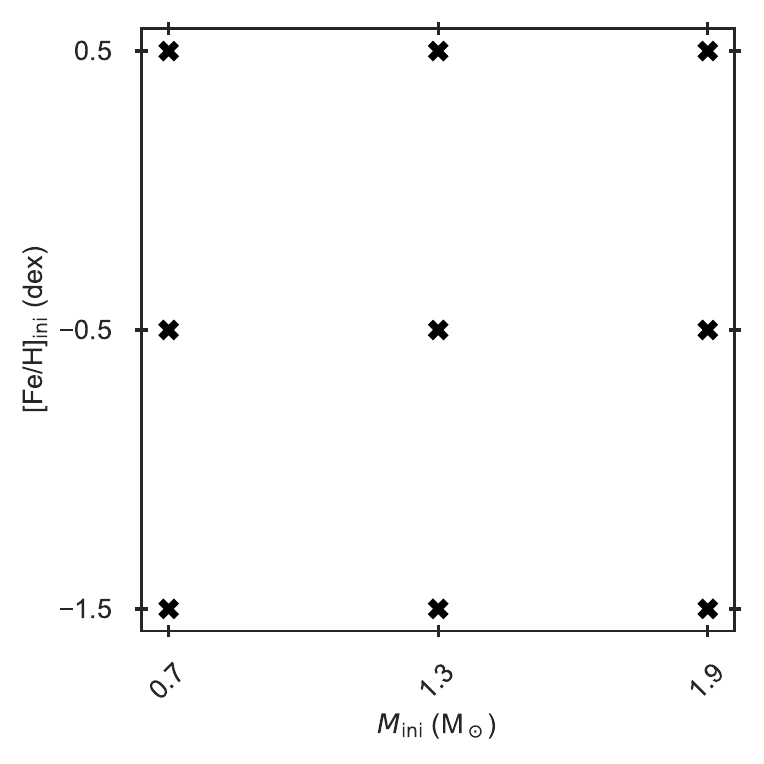}}
    \centering
    \caption{A simple plot showing the spanned parameter space of the validation grid in the two varied parameters, mass and $[\mathrm{Fe}/\mathrm{H}]$.}
    \label{fig:grid_coverage}
\end{figure}
All stellar evolution models employed in this work are computed using the Garching Stellar Evolution Code, GARSTEC \citep{Achim08}. The choice for equations of state are the ones from OPAL (\citealt{Rogers96}; \citealt{Rogers02}) and the work of Mihalas-Hummer-Däppen (\citealt{Däppen88}; \citealt{Hummer88}; \citealt{Mihalas88}; \citealt{Mihalas90}). We treat atomic diffusion following the prescription by \citet{Thoul94}. Different choices for opacities are used. For high temperatures we employ the ones from OPAL  (\citealt{Rogers92}; \citealt{Iglesias96}), and for low temperatures the opacities from \citet{Ferguson05}. The nuclear reaction rate cross-sections are from NACRE \citep{Angulo99}, with the exception of the reactions $^{14}\mathrm{N}(\mathrm{p},\gamma)^{15}\mathrm{O}$ and  $^{12}\mathrm{C}(\alpha,\gamma)^{16}\mathrm{O}$ which are from \citet{Formicula04} and \citet{Hammer05}, respectively. The stellar abundances used are from \citet{Asplund09}. 

The coverage of the validation grid is shown in Fig.~\ref{fig:grid_coverage}, which illustrates the Cartesian grid of 9 tracks presented in Sect.~\ref{sec:Validation}. It was constructed to widely span the parameter space of red giants and therefore have tracks at extreme combinations of mass and metallicity. Each track was evolved until the He flash. The track that evolved the furthest reached a $\Delta\nu = 0.810 \ \mu$Hz, while the least evolved reached $\Delta\nu = 0.877 \ \mu$Hz. In both cases, this is far into the TRGB region that we consider.

\end{appendix}

\end{document}